\documentclass[a4paper,11pt]{article}
\usepackage{jheppub} 
\usepackage{autobreak}

\usepackage{amsmath,amsfonts,amssymb,amscd}
\usepackage{graphicx}
\usepackage{mathrsfs}
\usepackage{bbm}
\usepackage{units}
\usepackage{xcolor}
\usepackage{tikz}
\usepackage{feynmp-auto}
\hypersetup{urlcolor=black}

\renewcommand{\arraystretch}{1.25}

\newcommand{\SO}[1]{\ensuremath{\mathrm{SO}(#1)}}
\newcommand{\SU}[1]{\ensuremath{\mathrm{SU}(#1)}}

\newcommand{\U}[1]{\ensuremath{\mathrm{U}(#1)}}

\newcommand{\rep}[2][]{\ensuremath{\boldsymbol{#2}#1}}

\renewcommand{\bar}[1]{\overline{#1}}

\definecolor{darkgreen}{HTML}{109930}

\title{Custodial Naturalness}

\author[a]{Thede de Boer,}
\author[a]{Manfred Lindner}
\author[a,b]{and Andreas Trautner}
\affiliation[a]{Max-Planck-Institut f\"ur Kernphysik, Saupfercheckweg 1, 69117 Heidelberg, Germany}
\affiliation[b]{CFTP, Departamento de F\'isica, Instituto Superior T\'ecnico, Universidade de Lisboa, Avenida Rovisco Pais 1, 1049 Lisboa, Portugal}

\emailAdd{thede.deboer@mpi-hd.mpg.de}
\emailAdd{lindner@mpi-hd.mpg.de}
\emailAdd{trautner@mpi-hd.mpg.de}

\abstract{
Custodial Naturalness is a new symmetry-based idea to explain the large separation between the electroweak~(EW) scale and ultraviolet completions of the Standard Model~(SM). Classical scale invariance is combined with an enhanced scalar-sector custodial symmetry and both are spontaneously broken by dimensional transmutation at a new intermediate scale. The SM-like Higgs boson is an elementary pseudo-Nambu-Goldstone-Boson~(pNGB) of the extended custodial symmetry, which naturally explains the suppression of the EW scale without a little hierarchy problem. We explain details of the general mechanism, its minimal realization and simplest extensions which populate Higgs-, gauge-, and neutrino portals and introduce candidates for particle Dark Matter~(DM). We show the stability of the mechanism under inclusion of new sources of explicit custodial symmetry violation, as well as under variations of boundary conditions at the high scale. Custodial Naturalness is experimentally testable -- including a specific correlation between the Higgs and top quark masses, as well as by the prediction of a new heavy $Z'$ gauge boson and a new dilaton-like scalar which are well-motivated targets for future colliders and Higgs factories. The cosmological evolution features a strongly supercooled phase transition implying that consequences of Custodial Naturalness may also be tested by gravitational wave observatories.
}

\begin{document}
\maketitle
\flushbottom

\widowpenalty10000\clubpenalty10000
\section{Introduction}
\label{sec:intro}
Remarkably, the Standard Model (SM) exhibits scale invariance at the classical level, explicitly broken only by the Higgs mass term as well as by quantum corrections.
The Higgs mass does not receive quadratically divergent corrections if there are no additional terms that break scale invariance~\cite{Bardeen:1995kv}.
Additional scales larger than the electroweak (EW) scale lead to corrections to the Higgs mass, giving rise to the hierarchy problem.

The anomalous breaking of classical scale invariance\footnote{In this work, we use the term ``conformal symmetry'' and classical scale invariance interchangeably.} can be translated to a physical scale via dimensional transmutation. As shown by Coleman and Weinberg \cite{Coleman:1973jx}, in the weak coupling regime, a massless scalar field can obtain a vacuum expectation value (VEV), if the beta function of the scalar quartic coupling is dominated by bosonic contributions. In the SM, the top Yukawa coupling dominates the running of the Higgs quartic coupling and thus the minimal realization, i.e.\ the SM without the Higgs mass term, is excluded~\cite{Weinberg:1976pe,Gildener:1976ih}. In models with additional scalar fields, dimensional transmutation in the new sector can generate an intermediate scale and the portal coupling then induces the Higgs mass (see, for example, Refs.~\cite{Hempfling:1996ht, Meissner:2006zh,Espinosa:2007qk,Chang:2007ki,Foot:2007as,Foot:2007ay,Iso:2009ss,Iso:2009nw,Oda:2015gna,Das:2015nwk,Das:2016zue}). Quantum corrections involving new fields with masses of the intermediate scale contribute to the Higgs mass giving rise to the little hierarchy problem. These contributions scale as $\sim\frac{1}{16\pi^2}g^2m_\psi^2$ where $m_\psi$ is the mass of the heavy field and $g$ is the coupling of the heavy field to the Higgs field. 

A popular approach to the little hierarchy problem is based on spontaneously broken approximate symmetries. The Higgs boson is then a pseudo-Nambu-Goldstone-Boson (pNGB) with a mass naturally smaller than the scale of new physics. Popular examples are strong coupling solutions such as composite Higgs~\cite{Kaplan:1983fs,Kaplan:1983sm,Georgi:1984af,Dugan:1984hq} and little Higgs~\cite{Arkani-Hamed:2001nha,Arkani-Hamed:2002sdy,Arkani-Hamed:2002ikv} as well as twin Higgs models~\cite{Chacko:2005pe,Barbieri:2005ri,Chacko:2005vw}. However, even minimal realizations of strong coupling solutions~\cite{Agashe:2004rs, Cacciapaglia:2014uja} or models that include conformal and/or custodial symmetry~\cite{Katz:2005au,Galloway:2010bp,Houtz:2016jxk,Ahmed:2023qsm},
typically require the introduction of one or several top partners at the $\mathrm{TeV}$ scale.

Recently, we proposed Custodial Naturalness~\cite{deBoer:2024jne} as a mechanism to explain both the separation of the EW and Planck scale as well as the little hierarchy problem. The simple setup of models with scale invariance and an elementary Higgs boson is combined with the idea of the Higgs boson as a pNGB of a spontaneously broken global symmetry. 
The top Yukawa coupling is marginal, similar to the SM, and can very well be present in the ultraviolet (UV) theory. Top partners are not required. While the top Yukawa coupling violates the enhanced global symmetry explicitly, it does not necessarily lead to a large correction to the Higgs mass for the same reasons as discussed by Bardeen \cite{Bardeen:1995kv}. Previous works have considered an elementary Higgs boson as a pNGB~\cite{Alanne:2014kea,Gertov:2015xma} compared to which we include scale invariance and dimensional transmutation, use a simpler scalar sector and impose an extended custodial symmetry at the high scale. Particle content and fermion charge assignment in the simplest realization of Custodial Naturalness closely resemble the ``minimal $B-L$ model''~\cite{Davidson:1978pm,Marshak:1979fm,Mohapatra:1980qe,Wetterich:1981bx,Jenkins:1987ue,Buchmuller:1991ce,Khalil:2006yi} and its conformal realization~\cite{Iso:2009ss,Iso:2009nw}, while the details of the scalar sector differ. 

In this work we extend the results of Ref.~\cite{deBoer:2024jne}, highlighting different aspects of custodial symmetry violation and their impact on the little hierarchy. We extend the original minimal model to incorporate new sources of custodial symmetry violation opening the possibility for the new sector to populate the neutrino mass matrix or allow for Dark Matter (DM). In total, we discuss three models - the minimal model, the neutrino portal model and the DM model.

After symmetry breaking, the particle content includes the SM in addition to a new gauge boson $Z'$ in the $\sim4-100\,\mathrm{TeV}$ range as well as the dilaton with a mass that is loop suppressed compared to the $Z'$ mass and typically in the $\sim30-1000\,\mathrm{GeV}$ range. The neutrino portal extension of the minimal model introduces heavy and massless new fermions while the DM model introduces a fermionic two-component DM candidate.

This work is structured as follows:
Section~\ref{sec. General discussion} presents the concept of Custodial Naturalness including an analytical discussion of the effective potential. Special emphasis is placed on the different sources of custodial symmetry violation. In Sec.~\ref{sec. models} we introduce different models that realize our idea. The effect of custodial symmetry violation is studied numerically and the amount of fine tuning is quantified. In Sec.~\ref{sec. constraints and signatures} we give numerical results for the masses of new particles and discuss experimental signatures of our model. Section~\ref{sec. finite T} sketches the thermal history of the Universe, and in Sec.~\ref{sec. future directions} we discuss variations of the general idea and embeddings. In Sec.~\ref{sec. Conclusion} we draw our conclusions.

\section{General discussion}\label{sec. General discussion}
The concept of Custodial Naturalness combines conformal and custodial symmetry in an enlarged scalar sector consisting of the SM Higgs doublet $H$ and an additional complex scalar singlet $\Phi$. The field $\Phi$ obtains a VEV spontaneously breaking custodial symmetry and the Higgs field is a pNGB associated with this breaking.
Before studying the scale invariant case, we briefly discuss the scalar potential with tree level mass terms. This allows us to understand how the mass of the Higgs field is protected by custodial symmetry.

\subsection{Non-conformal case}\label{sec. non-conformal}
The general tree level potential is given by
\begin{equation}
	V=-m_H^2|H|^2-m_\Phi^2|\Phi|^2+\lambda_H|H|^4+2\lambda_p|H|^2|\Phi|^2+\lambda_\Phi|\Phi|^4.
\end{equation}
Given that $m_\Phi^2>0$ and $-m_H^2+m_\Phi^2\frac{\lambda_p}{\lambda_\Phi}>0$, the minimum of the potential at tree level is given by $\langle\Phi\rangle:=\frac{v_\Phi}{\sqrt{2}}=\sqrt{\frac{m_\Phi^2}{2\lambda_\Phi}},\langle H\rangle=0$. We now integrate out the field corresponding to excitation in the $\Phi$ direction at tree level. This gives the potential in the effective field theory for $H$ given by
\begin{equation}
\begin{split}
	V_\text{EFT}=&\left(-m_H^2+\lambda_pv_\Phi^2\right)|H|^2+\left(\lambda_H-\frac{\lambda_p^2}{\lambda_\Phi}\right)|H|^4\\
	=&\left(-m_H^2+\lambda_p\frac{m_\Phi^2}{\lambda_\Phi}\right)|H|^2+\left(\lambda_H-\frac{\lambda_p^2}{\lambda_\Phi}\right)|H|^4.
\end{split}
\end{equation}
At tree level, the mass of $H$ vanishes for $m_H^2=m_\Phi^2$ and $\lambda_p=\lambda_\Phi$. Note how this is independent of the value of the coupling $\lambda_H$. If we now consider a potential that obeys  an approximate symmetry of rotations between $H$ and $\Phi$, where only the quartic coupling for $H$ breaks this symmetry, i.e.\
\begin{equation}
	V=-m^2\left(|H|^2+|\Phi|^2\right)+\lambda\left(|H|^2+|\Phi|^2\right)^2+(\lambda_H-\lambda)|H|^4\label{eq. V symm},
\end{equation}
then $H$ remains massless at tree level.
We have checked that $\langle\Phi\rangle=\sqrt{\frac{m_\Phi^2}{2\lambda_\Phi}},\langle H\rangle=0$ remains the minimum at tree level given that $\lambda_H>\lambda$. Generally speaking, symmetry violating interactions that only couple to the Higgs field (and not to $\Phi$) contribute to the Higgs mass only at subleading level. 

\subsection{Scalar sector and symmetry breaking}
Custodial Naturalness is based on the assumption that at some high scale, which we choose to be the Planck scale $M_\mathrm{Pl}$, the potential is scale invariant and has a $\mathrm{SO}(6)$ custodial symmetry,\footnote{We refer to this symmetry as custodial symmetry because it is a symmetry of the scalar potential that is explicitly broken by the gauge and Yukawa interactions~\cite{Sikivie:1980hm}. Under \SO{6}, the six scalar degrees of freedom transform as a real \rep{6}-plet.} explicitly
\begin{equation}
	V=\lambda\left(|H|^2+|\Phi|^2\right)^2\qquad\text{at}\qquad\mu=M_{\mathrm{Pl}}\;.
\end{equation}
 Classical scale invariance, which forbids the tree level mass terms, is broken by the scale anomaly, i.e.\ the non-vanishing beta functions.
\begin{figure}
	\centering
	\includegraphics[width=0.5\textwidth]{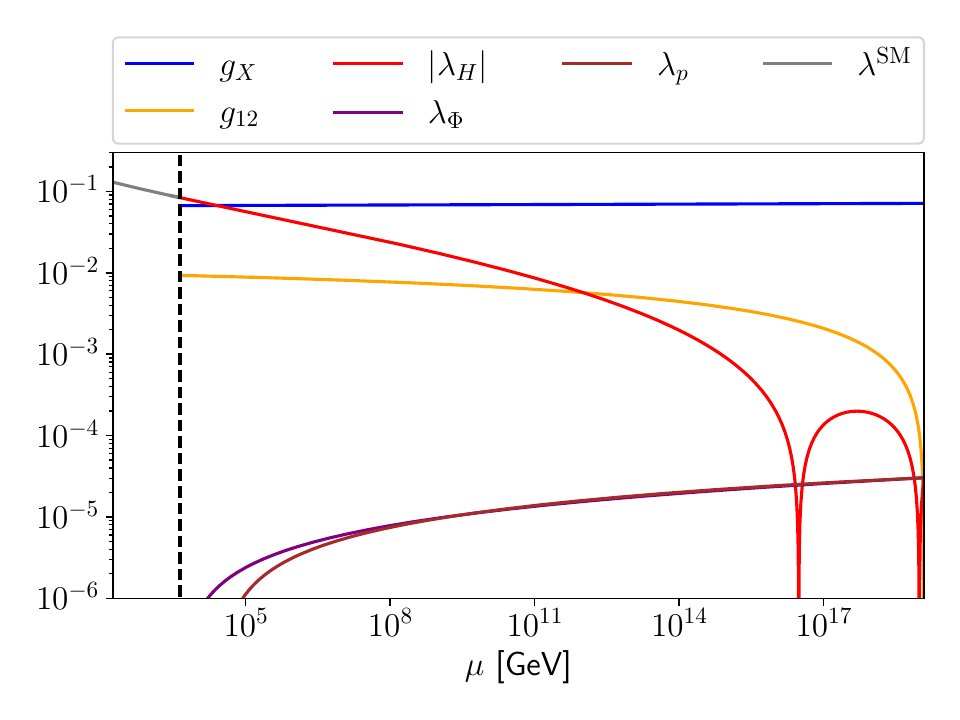}%
    \includegraphics[width=0.5\textwidth]{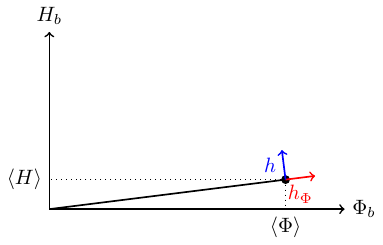}
	\caption{Left: Running of couplings for a typical model. At Planck scale, the scalar potential has a \SO{6} custodial symmetry $\lambda_H=\lambda_p=\lambda_\Phi$. The dashed vertical line indicates the scale of Coleman-Weinberg symmetry breaking. Right: The orientation of the VEV in the $\Phi-H$ plane and the radial excitation corresponding to the dilaton $h_\Phi$ as well as the orthogonal excitation corresponding to the  pNGB Higgs boson $h$.}
	\label{fig. running}
\end{figure}
Custodial symmetry breaking is mediated to the potential by quantum corrections as manifest in the RG running (see Fig.~\ref{fig. running}) demanding a more general form of the tree level quartic potential,
\begin{equation}
	V_\text{tree}=\lambda_H|H|^4+2\lambda_p|H|^2|\Phi|^2+\lambda_\Phi|\Phi|^4.
\end{equation}	
Nonetheless, $\lambda_\Phi$ and $\lambda_p$ remain close to each other as the difference $\lambda_p-\lambda_\Phi$ is protected by custodial symmetry, while $\lambda_H$ is driven to large positive values by the top Yukawa coupling. At some intermediate scale of $\sim 10^5\,\mathrm{GeV}$, $\lambda_p$ turns negative and $\lambda_\Phi$ turns small. At this scale the one loop potential needs to be considered in order to obtain the vacuum configuration~\cite{Coleman:1973jx}.

To study the structure of the VEVs, we first use the Gildener-Weinberg approximation~\cite{Gildener:1976ih}.
Adopting the typical notation, we write the  scale invariant potential as 
\begin{equation}
    V=f_{ijkl}\Phi_i\Phi_j\Phi_k\Phi_l,
\end{equation}
where $\Phi_i$ denotes the scalar fields\footnote{$\Phi_i$ takes values of all scalar fields and should not be confused with the scalar singlet $\Phi$.} and $f_{ijkl}$ are the corresponding quartic couplings.
At some RG scale $\mu_\mathrm{GW}$ the potential develops a flat direction. We write the fields in terms of the radial distance from the origin $\phi$ and a unit vector $n_i$ by $\Phi_i=n_i\phi$. The condition for the flat direction is then given by
\begin{equation}
    \left.\frac{\partial V}{\partial\Phi_j}\right|_{\Phi_i=n_i\phi}=0\qquad\text{and}\qquad\left. V\right|_{\Phi_i=n_i\phi}=0\quad\text{at }\mu=\mu_{\mathrm{GW}}
\end{equation}
for a non zero value of $\phi$. For negative $\lambda_p$, the solution to these equations is given by~\cite{Sher:1988mj}
\begin{equation}
    H=\sqrt{\frac{-\lambda_p}{\lambda_H-\lambda_p}}\phi,\qquad\Phi=\sqrt{\frac{\lambda_H}{\lambda_H-\lambda_p}}\phi,\qquad\lambda_\Phi=\frac{\lambda_p^2}{\lambda_H}.\label{eq. GW solution}
\end{equation}
Since $\lambda_p$ turns out to be small, the flat direction is mostly aligned in the $\Phi$-direction (see Fig.~\ref{fig. running}) similar to the scenario in Refs.~\cite{Kannike:2021iyh,Huitu:2022fcw,Kannike:2022pva}. 
Along this flat direction, quantum corrections generate a non-trivial minimum leading to a VEV that spontaneously breaks conformal and custodial symmetry. 
Custodial symmetry is broken like $\mathrm{SO}(6)\to\mathrm{SO}(5)$ which yields five Goldstone bosons, four of which are eaten by the longitudinal degrees of freedom of the massive gauge bosons. The final Goldstone boson is a pNGB that closely resembles the physical Higgs boson, whose mass is proportional to the size of custodial symmetry violation. The remaining massive scalar is the pNGB corresponding to the spontaneous breakdown of scale invariance. The mass of this dilaton, the radial excitation, is generated at one loop and therefore suppressed with respect to the VEV by the beta function.

\subsection{Charge assignment}
The particle content of our model consists of the SM fields in addition to three right-handed neutrinos $\nu_R$ and the complex scalar singlet $\Phi$. We further add a \U{1} gauge group whose contributions to the RGE drive $\lambda_\Phi$ to its critical value ensuring symmetry breaking \`{a} la Coleman-Weinberg. For minimal models, gauge anomaly freedom requires that the SM fermions have a \U{1} charge which is a linear combination of $B-L$ and hypercharge. We give the $B-L$ charges of the SM fields and $\Phi$ in Tab.~\ref{tab. particle content U1 model}. Additional fermions are vector-like and do not contribute to the gauge anomalies. It turns out to be convenient to work in a basis where the \U{1} charges of the scalar fields are symmetric. In this basis, the charges under the new $\U{1}_\mathrm{X}$ group are the following linear combination
\begin{equation}
	Q^{(\mathrm{X})}=2Q^{(\mathrm{Y})}+\frac{1}{q_\Phi}Q^{(\mathrm{B-L})},\label{eq. charges X}
\end{equation}
where $Q^{(\mathrm{Y})}$ and $Q^{(\mathrm{B-L})}$ are the hyper- and $B-L$ charges of a generic field, while $q_\Phi$ is the $B-L$ charge of $\Phi$ which is a free parameter of the charge assignment.
\begin{table}
	\centering
	\caption{Charges of SM and new fields under the new $\U{1}_\mathrm{X}$ gauge group. $B-L$ charges are linear combinations of $X$ and $Y$ charges and are shown as well. The ``Minimal particle content'' is sufficient to realize the idea of Custodial Naturalness. Also shown are minimal extensions that allow for new neutrino Yukawa couplings or an explanation of DM. Here, $q_\Phi$ and $p$ are free parameters of the charge assignment (see text for details). }
	\label{tab. particle content U1 model}
    \begingroup
    \renewcommand{\arraystretch}{1.5}
	\begin{tabular}{|cccc|c|}
		\hline
		Name & Generations  & \multicolumn{2}{c|}{$\SU{3}_\mathrm{c}\times\SU{2}_\mathrm{L}\times\U{1}_\mathrm{Y}\times\U{1}_\mathrm{X}$} & $\U{1}_\mathrm{B-L}$ \\ 
		\hline 
		\multicolumn{5}{c}{Minimal particle content}\\\hline
		$Q$      & $3$ & \hspace{1.25cm} $\left(\rep{3},\rep{2},+\frac16\right)$ & $+\frac{1}{3}+\frac{1}{3q_\Phi}$ & $+\frac{1}{3}$ \\
		$L$      & $3$ & \hspace{1.25cm} $\left(\rep{1},\rep{2},-\frac12\right)$ & $-1-\frac{1}{q_\Phi}$ & $-1$ \\
		$u_R$    & $3$ & \hspace{1.25cm} $\left(\rep{3},\rep{1},+\frac{2}{3}\right)$ & $+\frac{4}{3}+\frac{1}{3q_\Phi}$ & $+\frac{1}{3}$ \\
		$d_R$    & $3$ & \hspace{1.25cm} $\left(\rep{3},\rep{1},-\frac{1}{3}\right)$ & $-\frac{2}{3}+\frac{1}{3q_\Phi}$ & $+\frac{1}{3}$ \\
		$e_R$    & $3$ & \hspace{1.25cm} $\left(\rep{1},\rep{1},-1\right)$ & $-2-\frac{1}{q_\Phi}$ & $-1$ \\
		$\nu_R$  & $3$ & \hspace{1.25cm} $\left(\rep{1},\rep{1},\phantom{-}0\right)$ & $-\frac{1}{q_\Phi}$ & $-1$ \\\hline
		$H$      & $1$ & \hspace{1.25cm} $\left(\rep{1},\rep{2},+\frac12\right)$ & $+1$ & $\phantom{-}0$ \\
		$\Phi$   & $1$ & \hspace{1.25cm} $\left(\rep{1},\rep{1},\phantom{-}0\right)$ & $+1$ & $\phantom{-}q_\Phi$ \\\hline
		\multicolumn{5}{c}{Minimal set of additional fermions}\\\hline
		$\psi_L$ & $1$ & \hspace{1.25cm} $\left(\rep{1},\rep{1},0\right)$ & $-\left(\frac{1}{q_\Phi}+1\right)$ & $-(1+q_\Phi)$ \\
		$\psi_R$ & $1$ & \hspace{1.25cm} $\left(\rep{1},\rep{1},0\right)$ & $-\left(\frac{1}{q_\Phi}+1\right)$ & $-(1+q_\Phi)$ \\\hline
		\multicolumn{5}{c}{Additional fermions that allow for DM}\\\hline
		$\psi_L$ & $1$ & \hspace{1.25cm} $\left(\rep{1},\rep{1},0\right)$ & $\frac{p}{q_\Phi}$ & $p$ \\
		$\psi_R$ & $1$ & \hspace{1.25cm} $\left(\rep{1},\rep{1},0\right)$ & $\frac{p}{q_\Phi} +1$ & $p+q_\Phi$ \\
		$\psi'_L$& $1$ & \hspace{1.25cm} $\left(\rep{1},\rep{1},0\right)$ & $\frac{p}{q_\Phi} +1$ & $p+q_\Phi$ \\
		$\psi'_R$& $1$ & \hspace{1.25cm} $\left(\rep{1},\rep{1},0\right)$ & $\frac{p}{q_\Phi} $ & $p$ \\\hline
	\end{tabular}
    \endgroup
\end{table}

In this work, we discuss three different models. The minimal realization of Custodial Naturalness simply consists of the SM fields in addition to the scalar field $\Phi$. Further, we consider a model where we add an additional set of fermions, see Tab.~\ref{tab. particle content U1 model} (middle). The \U{1} charges of these new fermions are chosen in such a way that $\psi_L$ couples to right-handed neutrinos via the Yukawa interaction
\begin{equation}
	\mathcal{L}_\text{Yuk}\supset y_\psi^\alpha \bar{\psi}_L\Phi^\dagger\nu_R^\alpha+\mathrm{h.c.}\label{eq. Yukawa psi nuR}
\end{equation}
where $\alpha=1,2,3$ runs over the number of generations.
This model is the minimal model that allows for a Yukawa interaction involving $\Phi$.
The third model we discuss introduces two sets of new fermions (see Tab.~\ref{tab. particle content U1 model} (bottom)). Here, $p$ is a free parameter. If $p$ is set to a value that prohibits a Yukawa coupling involving new fermions and right-handed neutrinos, then the new fermions are stable, making them natural DM candidates. In this case we have the following new Yukawa interactions
\begin{equation}
	\mathcal{L}_\text{Yuk}\supset y_\psi \bar{\psi}_L\Phi^\dagger\psi_R+ y_{\psi'} \bar{\psi'}_L\Phi\psi'_R+\mathrm{h.c.}\label{eq. Yukawa psi psip}
\end{equation}

\subsection{Different sources of custodial symmetry violation}
The \SO{6} custodial symmetry is explicitly violated by the gauge and Yukawa interactions present in our model. In Sec.~\ref{sec. analytical discussion}, we will show that small values of the splitting $\lambda_\Phi-\lambda_p$ lead to a large hierarchy between the VEV of $\Phi$ and the Higgs mass. In contrast, $\lambda_H$ runs to large values driven by the top Yukawa coupling and only has a subleading effect on the hierarchy. This agrees with the result for the non-conformal case (Sec.~\ref{sec. non-conformal}) and might already be guessed from Eq.~(\ref{eq. GW solution}) in the conformal case. In order for the Higgs field to obtain a VEV, $\lambda_p<\lambda_\Phi$ at the intermediate scale is required (see Eq.~(\ref{eq. GW solution}) and Sec.~\ref{sec. analytical discussion}). 
We now discuss the different possible sources of custodial symmetry violation and how these contributions drive $\lambda_\Phi$ and $\lambda_p$ apart.

\subsubsection{SM sector}
The SM fermions and (electroweak) gauge bosons only couple to the Higgs field and have no coupling to $\Phi$. While these couplings strongly impact the running of $\lambda_H$, the effect on the running of $\lambda_p$ and $\lambda_\Phi$ is suppressed by $\lambda_p$. The difference in the beta functions of $\lambda_p$ and $\lambda_\Phi$ induced by SM couplings is given by
\begin{equation}
	\beta_{\lambda_p}-\beta_{\lambda_\Phi}\biggr|_\mathrm{SM}\simeq\frac{1}{16\pi^2}\lambda_p\left[-\frac92 g_L^2-\frac32 g_Y^2+12 \lambda_H +6 y_t^2\right],\label{eq. beta diff SM}
\end{equation}
where $g_Y$ and $g_L$ are the hypercharge and $\SU{2}_\mathrm{L}$ gauge couplings and $y_t$ is the top Yukawa coupling.
This tends to be a relatively small effect as for typical models $\lambda_p\lesssim 10^{-4}$. Such small values of $\lambda_p$ are required as, in order for $\lambda_\Phi$ to reach its critical value at $\mu_\mathrm{GW}$, the symmetric scalar coupling at the high scale needs to fulfill $\lambda\approx\frac{6g_X^4}{16\pi^2}\ln\left(\frac{M_\mathrm{Pl}}{\mu_\mathrm{GW}}\right)$. For $g_X\approx 0.1$ this requires $\lambda\approx10^{-4}$.  The SM contributions to $\beta_{\lambda_p}-\beta_{\lambda_\Phi}$ are negative for $10^{11}\,\mathrm{GeV}\lesssim\mu<M_{\rm Pl}$ and the integrated effect leads to $\lambda_p>\lambda_\Phi$. If this were the only source of custodial symmetry violation, the mass parameter for the Higgs doublet would be positive and thus there would be no EWSB. Consequently there need to be additional sources of custodial symmetry violation with opposite sign.\footnote{\label{footnote lower CS}An alternative might be to have a lower scale where custodial symmetry is realized. In this case the SM contribution might be sufficient to trigger EWSB (see also Sec.~\ref{sec. future directions}). }

\subsubsection{New gauge sector}
\begin{figure}
	\centering
	\includegraphics[width=\textwidth]{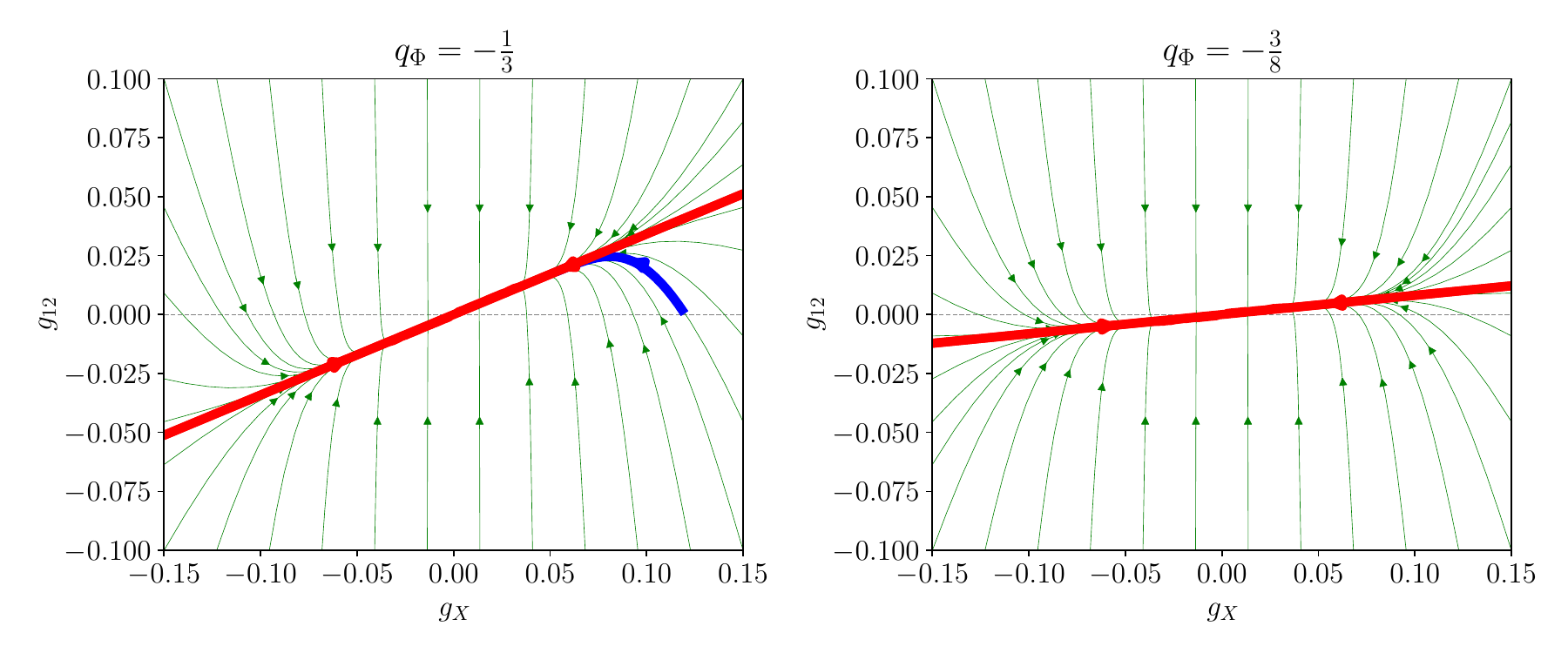}
	\caption{The trajectories of the RG flow in the $g_X-g_{12}$ plane calculated at one loop. The arrows represent the flow from the UV to the infrared (IR). $g_Y$ has been set to 0.48. The red lines correspond to $g_{12}=\frac{14}{41}g_X$ for $q_\Phi=-\frac13$ and $g_{12}=\frac{10}{123}g_X$ for $q_\Phi=-\frac38$. The blue line highlights a typical trajectory in the minimal model.}
	\label{fig. stream plot}
\end{figure}

Another source of custodial symmetry violation can be the couplings of the scalar fields to the $\U{1}_\mathrm{X}$ gauge boson. We work in the $\U{1}_{\mathrm{X}}$ basis where the charges of $\Phi$ and $H$ are equal. Changing the \U{1} basis shifts the custodial symmetry violation into the gauge kinetic mixing parameter.
This happens as the basis change modifies the covariant derivative. In the $B-L$ basis, with a gauge kinetic mixing parameter $\tilde g$\,\footnote{Gauge kinetic mixing is often introduced in the kinetic term as $\varepsilon F^{\mu\nu}F'_{\mu\nu}$~\cite{Galison:1983pa,Holdom:1985ag}. Through basis changes this term can be absorbed into a triangular gauge coupling matrix and the off-diagonal entry is given by $\tilde g:=\varepsilon g_Y /\sqrt{1-\varepsilon^2}$~\cite{Pan:2018dmu}.} and the new gauge coupling $g_{\mathrm{B-L}}$, the covariant derivative, restricted to \U{1} gauge bosons, acting on a generic field $\phi$ is given by
\begin{equation}
  	\left[\partial_\mu+i \left(Q^{(\mathrm{Y})},Q^{(\mathrm{B-L})}\right)\left(\begin{matrix}g_Y & \tilde g\\ 0 & g_{\mathrm{B-L}}\end{matrix}\right)\left(\begin{matrix}A_\mu^{(\mathrm{Y})}\\A_\mu^{(\mathrm{X})}\end{matrix}\right)\right]\phi.
\end{equation}
$A_\mu^{(\mathrm{Y})}$ and $A_\mu^{(\mathrm{X})}$ are the \U{1} gauge fields. Rewriting this in terms of the $\U{1}_{\mathrm{X}}$ charge $Q^{(\mathrm{X})}$ defined in Eq.~(\ref{eq. charges X}) yields
\begin{equation}
  	\left[\partial_\mu+i \left(Q^{(\mathrm{Y})},Q^{(\mathrm{X})}\right)\left(\begin{matrix}g_Y & \tilde g-2 q_\Phi g_{\mathrm{B-L}}\\ 0 & q_\Phi g_{\mathrm{B-L}}\end{matrix}\right)\left(\begin{matrix}A_\mu^{(\mathrm{Y})}\\A_\mu^{(\mathrm{X})}\end{matrix}\right)\right]\phi.
\end{equation}
We define the kinetic mixing parameter in the new basis as $g_{12}:=\tilde g-2 q_\Phi g_{\mathrm{B-L}}$ and the gauge coupling $g_X:=q_\Phi g_{\mathrm{B-L}}$.
In the presence of fields charged under both \U{1} groups, gauge kinetic mixing will generally be generated at the loop level \cite{Holdom:1985ag} and is therefore non-zero.
The difference in the beta functions of $\lambda_p$ and $\lambda_\Phi$ induced by gauge kinetic mixing is given by 
\begin{equation}
	\beta_{\lambda_p}-\beta_{\lambda_\Phi}\biggr|_{g_{12}}\simeq\frac{g_{12}}{16\pi^2}\left[6g_X^3 +\frac{3}{2}g_{12}g_X^2\right].\label{eq. delta beta gauge kin}
\end{equation}
In order to ensure that the splitting of $\lambda_\Phi$ and $\lambda_p$ does not become too large, $g_{12}$ needs to remains small under the RG flow. We find that the choice $q_\Phi=-\frac{16}{41}$ ensures that $g_{12}$ remains zero at one loop if it is set to zero at some initial scale.\footnote{The value $q_\Phi=-\frac{16}{41}$ was also found in~\cite{Oda:2015gna,Das:2016zue}.} This corresponds to the ``charge orthogonality condition'' \cite{Loinaz:1999qh} (see also~\cite{Carone:1995pu,Chang:2000xy,Aranda:2000ma}). In this work we will consider the cases $q_\Phi=-\frac13$ and $q_\Phi=-\frac38$.\footnote{Depending on the choice of $q_{\Phi}$, integer normalized charges may appear more or less appealing. This might not just be an aesthetical question but can be important for further model building and specifically the embedding of the $\U1_{\mathrm{X}}$ together with the other gauge groups into a larger structure.}
For values of $|q_\Phi|$ outside of roughly $|q_\Phi|\in\left[\frac13,\frac{5}{11}\right]$, $g_{12}$ runs to large values which leads to large contributions in Eq.~(\ref{eq. delta beta gauge kin}) spoiling custodial symmetry. In Fig.~\ref{fig. stream plot}, we show the flow of $g_{12}$ for the two values of $q_\Phi=-\frac13$~(left) and $q_\Phi=-\frac38$~(right).
In the left plot, $g_{12}$ converges to $g_{12}=\frac{14}{41}g_X$ and in the right plot to  $g_{12}=\frac{10}{123}g_X$ (at one loop). In the left plot, we also highlight a trajectory that starts at $g_{12}=0$ which corresponds to the setup in Sec.~\ref{sec. minimal model}.\footnote{We stress that the entire analysis of this paper can be done in the $B-L$ basis as long as $\tilde g$ is close to $-\frac23 g_\mathrm{B-L}$ ($-\frac34 g_\mathrm{B-L}$) for $q_\Phi=-\frac13(-\frac38)$ which is equivalent to our model with small $g_{12}$.}

For $|g_{12}|\lesssim |g_X|$, the first term in Eq.~(\ref{eq. delta beta gauge kin}) is dominant and the sign of the contribution depends on the relative sign of $g_{12}$ and $g_X$. If they have the same sign, then Eq.~(\ref{eq. delta beta gauge kin}) leads to $\lambda_p<\lambda_\Phi$ which is required for EWSB.

\subsubsection{New Yukawa interactions}
If the model includes new fermions with Yukawa couplings to $\Phi$, then these will contribute to $\beta_{\lambda_p}-\beta_{\lambda_\Phi}$ as
\begin{equation}
	\beta_{\lambda_p}-\beta_{\lambda_\Phi}\biggr|_{y_\psi}\simeq\frac{\sum_k 2y_{\psi_k}^4}{16\pi^2},\label{eq. beta yuk}
\end{equation}
where we assume real Yukawa couplings.
For the model with one new set of fermions the Yukawa interactions are given by Eq.~(\ref{eq. Yukawa psi nuR}) and the sum runs over the single value $\bar y_\psi:=\sqrt{y_\psi^\alpha y_\psi^\alpha}$. In the DM model, the Yukawa interactions are given by Eq.~(\ref{eq. Yukawa psi psip}) and the sum runs over $y_\psi$ and $y_{\psi'}$.
Such new Yukawa couplings lead to $\lambda_p<\lambda_\Phi$ which is required for EWSB. In the remainder of this work, the sum over $k$ is always understood as running over all Yukawa interaction involving $\Phi$.

\subsection{The effective potential}\label{sec. analytical discussion}
The Coleman-Weinberg potential for background fields $\Phi_b$ and $H_b$ at one loop in $\bar{\mathrm{MS}}$ is given by
\begin{equation}
	V_\text{eff}=V_\text{tree}+ \sum_i \frac{n_i (-1)^{2 s_i}}{64\pi^2}m_{i,\text{eff}}^4\left[\ln\left(\frac{m_{i,\text{eff}}^2}{\mu^2}\right)-C_i\right],\label{eq. V_eff}
\end{equation}
where $n_i$ is the number of degrees of freedom for the corresponding field, $(-1)^{2 s_i}$ is $+1$ for bosons and $-1$ for fermions. $C_i$ is given by $\frac32$ for fermions and scalar fields and $\frac56$ for vector bosons.
The sum runs over all fields present in the respective version of our model. The effective masses $m_{i,\text{eff}}$ for the neutral gauge bosons are given by the eigenvalues of
\begin{equation}
\begin{split}
	M_V=\left(\begin{matrix}
	\frac{g_Y^2}{2} H_b^2 &-\frac{g_Y g_L}{2} H_b^2 & \frac{\left( 2g_X+g_{12}\right)g_Y}{2} H_b^2\\
	-\frac{g_Y g_L}{2}H_b^2 & \frac{g_L^2}{2} H_b^2 &-\frac{\left( 2g_X+g_{12}\right)g_L}{2}H_b^2 \\
	\frac{\left( 2g_X+g_{12}\right)g_Y}{2} H_b^2 &-\frac{\left( 2g_X+g_{12}\right)g_L}{2} H_b^2 & 2\left(\frac{2g_X+g_{12}}{2}\right)^2 H_b^2+2g_X^2\Phi_b^2
	\end{matrix}\right).\label{eq. vector effective masses}
 \end{split}
\end{equation}
For charged gauge bosons we have $m_{W^\pm,\text{eff}}^2=\frac{g_L^2}{2}H_b^2$ and for the top quark $m_{t,\text{eff}}=y_tH_b$. The effect of the other SM fermions can be neglected due to their small Yukawa couplings.
The scalar effective masses, determined by the second derivative of the tree level potential, are given by $\left(2\lambda_p \Phi_b^2 + 2 \lambda_H H_b^2,2\lambda_p \Phi_b^2 + 2 \lambda_H H_b^2,2\lambda_p \Phi_b^2 + 2 \lambda_H H_b^2,2\lambda_p H_b^2 + 2 \lambda_\Phi \Phi_b^2\right)$ and the eigenvalues of 
\begin{equation}
    \left(\begin{matrix}
        2\lambda_p H_b^2+6\lambda_\Phi\Phi_b^2 & 4 \lambda_p H_b \Phi_b \\
        4 \lambda_p H_b \Phi_b & 2\lambda_p \Phi_b^2+6\lambda_H H_b^2     \end{matrix}\right).
\end{equation}
For the model with one additional set of fermions, there is an additional effective mass given by $m^2_{\psi,\text{eff}}=y_\psi^\alpha y_\psi^\alpha\Phi_b^2$. In the DM model one needs to include the effective masses $m_{\psi,\text{eff}}=y_\psi\Phi_b$ and $m_{\psi',\text{eff}}=y_{\psi'}\Phi_b$.
The values of $\Phi_b$ and $H_b$ at the minimum of the Coleman-Weinberg potential are the VEVs $\langle\Phi\rangle$ and $\langle H\rangle$. 

Running from the custodially symmetric point at $M_\mathrm{Pl}$ down to the intermediate scale, the top Yukawa coupling drives $\lambda_H$ to large positive values. Therefore, the flat direction is mostly aligned with $\Phi$ such that $\langle\Phi\rangle\gg\langle H\rangle$ (see Eq.~(\ref{eq. GW solution})). In order to obtain insight into the effect of custodial symmetry violation on the Higgs potential, we will now discuss analytic approximations of the Coleman-Weinberg potential. To this end, we first solve the minimum condition for $\Phi_b$ for general $H_b$ and use this to implicitly define $\tilde\Phi(H_b)$ via
\begin{equation}
	\left.\frac{\partial V_\text{eff}}{\partial \Phi_b}\right|_{\Phi_b=\tilde{\Phi}(H_b)}=0.
\end{equation}
This procedure resembles Effective Field Theory (EFT) methods~\cite{Burgess:2007pt,Manohar:2020nzp} noting, however, that the CW potential is a function rather than a functional of the constant background fields.
The VEV of $\Phi$ in the limit $\langle\Phi\rangle\gg\langle H\rangle$ can be approximated by $\Phi_0:=\tilde\Phi(H_b=0)$ which is given by 
\begin{equation}
	\ln\left(\frac{\Phi_0^2}{\mu^2}\right)\!=\!-\frac{16 \pi^2 \lambda _{\Phi }\!+\!\left\{g_X^4 \left[3 \ln \left(2 g_X^2\right)-1\right]\!+\!4\lambda _p^2 \left[\ln \left(2\lambda _p\right)-1\right]\!-\!\sum_k \left[y_{\psi_k }^4 \left(\ln y_{\psi_k}^2-1\right)\right]\right\}}{ \left(3 g_X^4+4\lambda _p^2-\sum_k y_{\psi_k}^4\right)}.\label{eq. vev Phi_0}
\end{equation}
We use the same summation for the Yukawa couplings as in Eq.~(\ref{eq. beta yuk}). This is the usual result of dimensional transmutation. We now define a new potential for $H_b$ as
\begin{equation}
	V_\text{EFT}(H_b)~:=~V_\text{eff}(H_b,\tilde{\Phi}(H_b)).\label{eq. def VEFT}
\end{equation}
This new potential has a minimum at $H_b=\langle H\rangle$ since 
\begin{equation}
	\left.\frac{\partial V_\text{EFT}}{\partial H_b}\right|_{H_b=\langle H\rangle}=
	\left.\frac{\partial V_\text{eff}}{\partial H_b}+\frac{\partial V_\text{eff}}{\partial \Phi_b}\frac{\partial \tilde{\Phi}}{\partial H_b}\right|_{H_b=\langle H\rangle,\Phi_b=\langle\Phi\rangle}=0,
\end{equation}	
where the last equality is true since, by definition, $\langle\Phi\rangle$ and $\langle H\rangle$ fulfill the minimum conditions for $V_\text{eff}$. We now expand $V_\text{EFT}$ in orders of $H_b/\Phi_0$ and find at quadratic order
\begin{equation}
\begin{split}
    V_\text{EFT}\supset&2\left[\lambda_p-\frac{3\left(g_X+\frac{g_{12}}{2}\right)^2g_X^2}{3g_X^4+4\lambda_p^2-\sum_k y_{\psi_k}^4}\left(\lambda_\Phi+\sum_k\left\{\frac{y_{\psi_k}^4}{16\pi^2}\left[\frac{2}{3}+\ln\left(\frac{2g_X^2}{y_{\psi_k}^2}\right)\right]\right\}\right)\right]\Phi_0^2H_b^2\\
    &\hspace{6ex}+\frac{\lambda_p\lambda_H}{16\pi^2}\left[...\right]\Phi_0^2H_b^2,\raisetag{17pt}\label{eq. VEFT quadratic}
\end{split}
\end{equation}
where $[...]$ are $\mathcal{O}(1)$ terms suppressed by the $\lambda_p\lambda_H/(16\pi^2)$ prefactor.
This expression shows how the different sources of custodial symmetry violation impact the Higgs potential.
For small $y_\psi$ and $g_{12}$, the quadratic (mass) term for the Higgs field is $\approx 2(\lambda_p-\lambda_\Phi)\Phi_0^2H_b^2$. Note that the SM custodial symmetry violations (i.e.\ the top Yukawa and the electroweak gauge contributions) do not show up in this expression completely in line with our discussion in Sec.~\ref{sec. non-conformal}.

The RG scale $\mu$ should be chosen close to $\Phi_0$ in order to avoid large logarithms. It turns out that a particularly convenient choice is $\mu=\mu_0:=\sqrt2 g_X\Phi_0e^{-1/6}$. At this scale $|\lambda_\Phi|\ll|\lambda_p|$ and the quadratic term (i.e.\ Eq.~(\ref{eq. VEFT quadratic})) simplifies to 
\begin{equation}
	V_\text{EFT}\supset2\lambda_p\left\{1+\frac{[4\lambda_p+6\lambda_H]\left[\ln\left(\frac{2\lambda_p\Phi_0^2}{\mu_0^2}\right)-1\right]}{16\pi^2}\right\}H_b^2 \Phi_0^2.
\end{equation}	
Note how this quadratic term, which resembles the Higgs mass term, is $\approx 2\lambda_p\Phi_0^2H_b^2$ and therefore $\lambda_p\Phi_0^2$ should be of the order of the electroweak scale squared.
This agrees with the flat direction in the Gildener-Weinberg approximation, in the sense that Eq.~(\ref{eq. GW solution}) implies
\begin{equation}
	\lambda_H H^2=-\lambda_p\Phi^2
\end{equation}
along the flat direction. While for Eq.~(\ref{eq. VEFT quadratic}) we expanded in powers of $H_b/\Phi_0$, we now introduce a new artificial expansion parameter $\epsilon$ as
\begin{equation}
	\frac{H_b}{\Phi_0}\to\epsilon\frac{H_b}{\Phi_0},\qquad \lambda_p\to\epsilon^2\lambda_p.
\end{equation}
 Sending $\epsilon\to0$ corresponds to a 't Hooft-Veneziano-like limit \cite{tHooft:1973alw,Veneziano:1976wm}
\begin{equation}
	\frac{\Phi_0}{H_b}\to \infty,\qquad\frac{\lambda_p}{\lambda_H}\to0,\qquad\lambda_p \Phi_0^2=\lambda_H H_b^2\text{ (fixed)}.
\end{equation}
Expanding Eq.~(\ref{eq. def VEFT}) in powers of $\epsilon$ up to $\epsilon^4$ we find at $\mu=\mu_0$
\begin{equation}
\begin{split}
	V_\text{EFT}=&\frac{-3g_X^4+\sum_k y_{\psi_k}^4}{32\pi^2}\Phi_0^4+2\lambda_p\Phi_0^2H_b^2+\lambda_HH_b^4\\
    &+\sum_i\frac{n_i (-1)^{2s_i}}{64\pi^2}m_{i,\text{eff}}^4\left[\ln\left(\frac{m_{i,\text{eff}}^2}{\mu_0^2}\right)-C_i\right] -\frac{3\left(\frac{g_{12}}{2}+g_X\right)^4\left(\sum_k y_{\psi_k}^4\right)}{16\pi^2\left(3 g_X^4-\sum_k y_{\psi_k}^4\right)}H_b^4,\label{eq. V_EFT conf}
\end{split}
\end{equation}
where the sum over $i$ runs over the effective masses in the SM with a tree level potential of $V_\text{tree}=2\lambda_p\Phi_0^2H_b^2+\lambda_HH_b^4$. Eq.~(\ref{eq. V_EFT conf}) agrees with the SM effective potential at one loop~\cite{Martin:2001vx} up to the last term which gives a correction to the Higgs quartic coupling. We note that expanding Eq.~(\ref{eq. def VEFT}) in $\epsilon$ up to $\epsilon^4$ drops terms $\propto \lambda_p^3\Phi_0^4,\, \lambda_p^2 H_b^2\Phi_0^2,\, \lambda_pH_b^4$ when compared to expanding in $H_b/\Phi_0$ up to $\left(H_b/\Phi_0\right)^4$.

\subsection{Masses and mixing}
With the vacuum structure understood, we now turn to the masses of the new particles.
\subsubsection{Scalar masses}
We calculate the scalar mass matrix by taking the second derivatives of Eq.~(\ref{eq. V_eff}), i.e.\ $m^2_{ab}=\partial_{\phi_a}\partial_{\phi_b}V_\text{eff}$ evaluated at $H_b=\langle H \rangle=v_H/\sqrt{2}$ and $\Phi_b=\langle \Phi \rangle=v_\Phi/\sqrt{2}$. For the dilaton mass we find
\begin{equation}
    m_{h_\Phi}^2\approx\frac{3g_X^4-\sum_k y_{\psi_k}^4+4\lambda_p^2}{4\pi^2}\frac{v_\Phi^2}{2}\approx \beta_{\lambda_\Phi}v_\Phi^2\label{eq. dilaton mass}.
\end{equation}
The dilaton is the pNGB associated with spontaneous breaking of scale symmetry and the mass is suppressed by the scale anomaly, i.e.\ the beta function.
The mass of the physical Higgs boson is approximated by (at $\mu=\Phi_0$, $\lambda_p$ is smaller than zero)
\begin{equation}
	m_h^2\approx2\left[-\lambda_p+\frac{3\left(g_X+\frac{g_{12}}{2}\right)^2g_X^2}{3g_X^4+4\lambda_p^2-\sum_k y_{\psi_k}^4}\left(\lambda_\Phi+\sum_k\left\{\frac{y_{\psi_k}^4}{16\pi^2}\left[\frac{2}{3}+\ln\left(\frac{2g_X^2}{y_{\psi_k}^2}\right)\right]\right\}\right)\right]v_\Phi^2.
\end{equation}
In the limit of vanishing $g_{12}$ and $y_{\psi_k}$, this reduces to $m_h^2\approx2\left(\lambda_\Phi-\lambda_p\right)v_\Phi^2$. 
The physical Higgs boson is the pNGB associated with spontaneous breaking of \SO{6} custodial symmetry and the mass is generated via the differential running of $\lambda_p$ and $\lambda_\Phi$.
The Higgs-dilaton mixing angle is approximately given by
\begin{equation}
	\tan\theta\!\approx\! \frac{2\left[\lambda_p-\frac{3\left(g_X+\frac{g_{12}}{2}\right)^2g_X^2}{3g_X^4+4\lambda_p^2-\sum_ky_\psi^4}\left(\lambda_\Phi+\sum_k\left\{\frac{y_\psi^4}{16\pi^2}\left[\frac{2}{3}+\ln\left(\frac{2g_X^2}{y_\psi^2}\right)\right]\right\}\right)+\frac{3\left(g_X+\frac{g_{12}}{2}\right)^2g_X^2}{16\pi^2}\right]v_\Phi v_H}{m_h^2-m_{h_\Phi}^2}\label{eq. theta Higgs dilaton}
\end{equation}
and takes values of $\tan\theta\lesssim 10^{-2}$ for typical viable points. 
\subsubsection{Vector masses}
The mass matrix for the neutral gauge bosons $M_V$ is given by Eq.~(\ref{eq. vector effective masses}) evaluated at the VEVs $H_b=\langle H \rangle=v_H/\sqrt{2}$ and $\Phi_b=\langle \Phi \rangle=v_\Phi/\sqrt{2}$. The eigenvalues are obtained by $U^TM_VU$ with
\begin{equation}
    U=\left(\begin{matrix}c&-s c'&s s'\\s&c c'&-c s'\\0&s'&c'\end{matrix}\right),
\end{equation}
where $s=\sin(\theta_W)$ and $c=\cos(\theta_W)$ with the electroweak mixing angle $\theta_W=\arctan\left(\frac{g_Y}{g_L}\right)$ and $s'=\sin(\theta')$ and $c'=\cos(\theta')$ with
\begin{equation}
    \tan(2\theta')=-\frac{2(g_{12}+2g_X)\sqrt{g_L^2+g_Y^2}v_H^2}{\left[g_L^2+g_Y^2-(g_{12}+2g_X)^2\right]v_H^2-4g_X^2v_\Phi^2}.
\end{equation}
The masses of the $Z$ and $Z'$ bosons are given by 
\begin{align}\label{eq:mZmZp}
        m_Z^2=&\;\frac{1}{2}(g_L^2+g_Y^2)\frac{v_H^2}{2}\left[1-\frac{\left(\frac{g_{12}}{2}+g_X\right)^2}{g_X^2}\frac{v_H^2}{v_\Phi^2}+\mathcal{O}\left(\frac{v_H^4}{v_\Phi^4}\right)\right],\\
        m_{Z'}^2=&\; 2g_X^2\frac{v_\Phi^2}{2}+\frac12(g_{12}+2g_X)^2\frac{v_H^2}{2}+\mathcal{O}\left(\frac{v_H^4}{v_\Phi^2}\right),
\end{align}
while the photon remains massless.
The $Z$ mass is shifted compared to the SM prediction constraining the scale of CW symmetry breaking. However, we will see below that this constraint turns out to be weaker than the limits from direct $Z'$ searches.

\section{Different models realizing Custodial Naturalness}\label{sec. models}
\subsection{Minimal model}\label{sec. minimal model}
The minimal model that realizes Custodial Naturalness consists of the fields given in Tab.~\ref{tab. particle content U1 model} with no additional fermions. We choose $q_\Phi=-\frac13$ which is the same setup as in Ref.~\cite{deBoer:2024jne}. In this section we reproduce the main findings and give more details on how custodial symmetry violation in form of gauge kinetic mixing $g_{12}$ affects the hierarchy. 

A natural boundary condition for gauge kinetic mixing at $M_\mathrm{Pl}$ is $g_{12}\bigr|_{M_\mathrm{Pl}}=0$, which enhances custodial symmetry at the high scale. With this condition imposed, and taking $\lambda_H\bigr|_{M_\mathrm{Pl}}=\lambda_p\bigr|_{M_\mathrm{Pl}}=\lambda_\Phi\bigr|_{M_\mathrm{Pl}}$, the model has the same number of free parameters as the SM. Along the RG flow, $g_{12}$ runs towards non-zero values (see Fig.~\ref{fig. stream plot}) guaranteeing $\lambda_p-\lambda_\Phi<0$ and therefore EWSB. The equality of quartic couplings by custodial symmetry is an assumption that we expect to be explained by some mechanism at the high scale. Since we already know that this symmetry is only approximate, we stress that none of our mathematical results, e.g.\ absence of fine tuning in mass hierarchies and pNGB nature of the Higgs boson, are affected by small variations in the boundary conditions even if they slightly break custodial symmetry. As a proxy for such variations, we will allow for non-zero $g_{12}$ at Planck scale to open up the parameter space. 

We explore the parameter space by means of a random scan. In order to find reasonable starting points at $M_\mathrm{Pl}$, we sample the input values at the low scale and run these couplings up to the Planck scale where we impose custodial symmetry $\lambda_\Phi\bigr|_{M_\mathrm{Pl}}=\lambda_p\bigr|_{M_\mathrm{Pl}}=\lambda_H\bigr|_{M_\mathrm{Pl}}$. This set of parameters is then run down to $\mu_0$ where we calculate the VEVs and scalar masses before matching to the SM. 

More precisely, we choose the top pole mass in the $3\sigma$ range $M_t\in [170.4,174.6]\,\mathrm{GeV}$. The $\overline{\text{MS}}$ values for the SM gauge and Yukawa couplings are then obtained using the formulae in Ref.~\cite{Buttazzo:2013uya}, while the parameters in the SM Higgs potential (i.e.\ $\lambda_H^\mathrm{SM}$ and $m_H^\mathrm{SM}$) are chosen in such a way that the one loop effective potential reproduces the central values of the Higgs VEV and mass at $\mu=M_t$. We then run all couplings up to a randomly chosen scale $\tilde\mu_0\in\left[500,10^6\right]\,\mathrm{GeV}$ using the SM two loop RGEs and choose a random value for $g_X\bigr|_{\tilde\mu_0}\in[0,0.20]$. Eq.~(\ref{eq. vev Phi_0}) together with $\tilde\mu_0=\sqrt{2}g_X\Phi_0 e^{-1/6}$ allows us to derive $\lambda_\Phi\bigr|_{\tilde\mu_0}$. We also set $\lambda_H\bigr|_{\tilde\mu_0}=\lambda_H^\mathrm{SM}\bigr|_{\tilde\mu_0}$ and $\lambda_p\bigr|_{\tilde\mu_0}=\lambda_\Phi\bigr|_{\tilde\mu_0}$ which are reasonable estimates and the precise values will be set by custodial symmetry at the high scale. With all couplings fixed, we use the two loop RGEs obtained with \texttt{PyR@TE}~\cite{Sartore:2020gou} to run up to $M_\mathrm{Pl}$ where we impose \SO{6} custodial symmetry by the formal replacement $\left.\lambda_H,\lambda_p\right|_{M_\mathrm{Pl}}\rightarrow \lambda$ with $\lambda:=\lambda_\Phi\bigr|_{M_\mathrm{Pl}}$. We choose $g_{12}\bigr|_{M_\mathrm{Pl}}=0$ or $g_{12}\big|_{M_\mathrm{Pl}}\in[-0.1,0.1]\cdot g_X\bigr|_{M_\mathrm{Pl}}$ for the gauge kinetic mixing parameter. This defines a set of sensible starting parameters at $M_\mathrm{Pl}$.

Given a set of couplings at $M_\mathrm{Pl}$, we use the two loop RGEs to run down to a new scale $\mu_0$, which is found by iteratively using Eq.~(\ref{eq. vev Phi_0}) as well as the definition of $\mu_0$. At this scale, we calculate the VEVs and scalar masses numerically from the full effective potential Eq.~(\ref{eq. V_eff}). Matching to the SM is done by requiring that the SM effective potential at $\mu_0$ gives the same values for the electroweak VEV and Higgs mass as Eq.~(\ref{eq. V_eff}). This allows us to fix the couplings in the SM Higgs potential (i.e.\ $\lambda_H^\mathrm{SM}$ and $m_H^\mathrm{SM}$). These couplings are run down to $M_t$ where we calculate the Higgs VEV and mass from the SM effective potential and the top pole mass by inverting the formula in Ref.~\cite{Buttazzo:2013uya}. We exclude all points that yield a Higgs VEV outside of $v_H^\mathrm{exp}\pm0.1\,\mathrm{GeV}$.
This is the same setup as in Ref.~\cite{deBoer:2024jne} and, in case of the minimal model, also the same data is used.
Unless stated otherwise we do not impose a constraint on the Higgs mass. We have checked that the points with correct Higgs mass are roughly evenly distributed in the allowed parameter space.

\begin{figure}
	\centering
	\includegraphics[width=0.5\textwidth]{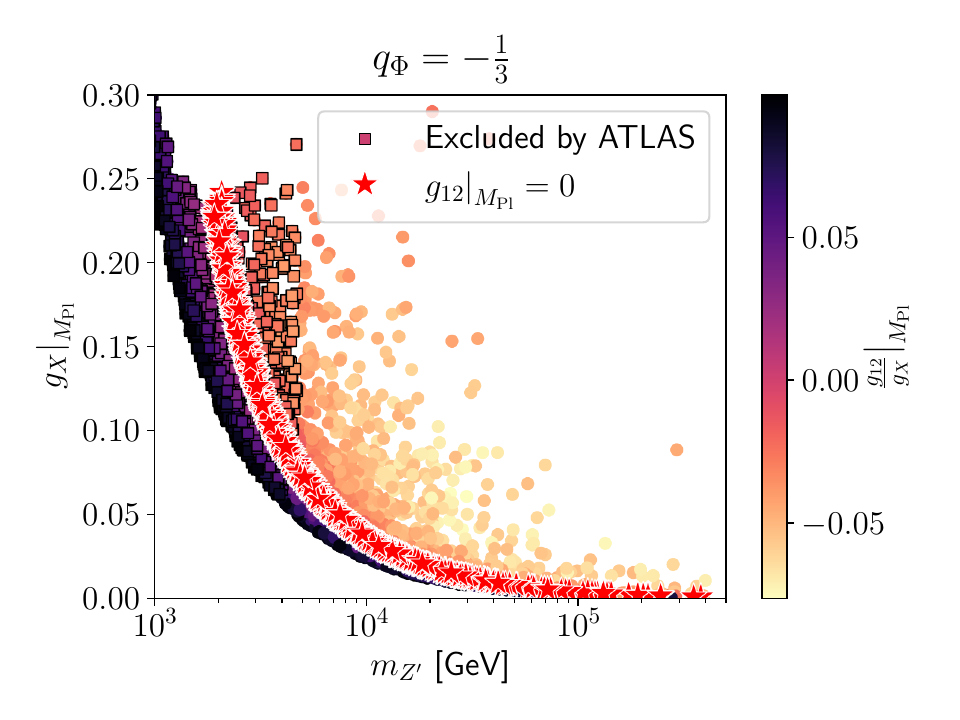}%
	\includegraphics[width=0.5\textwidth]{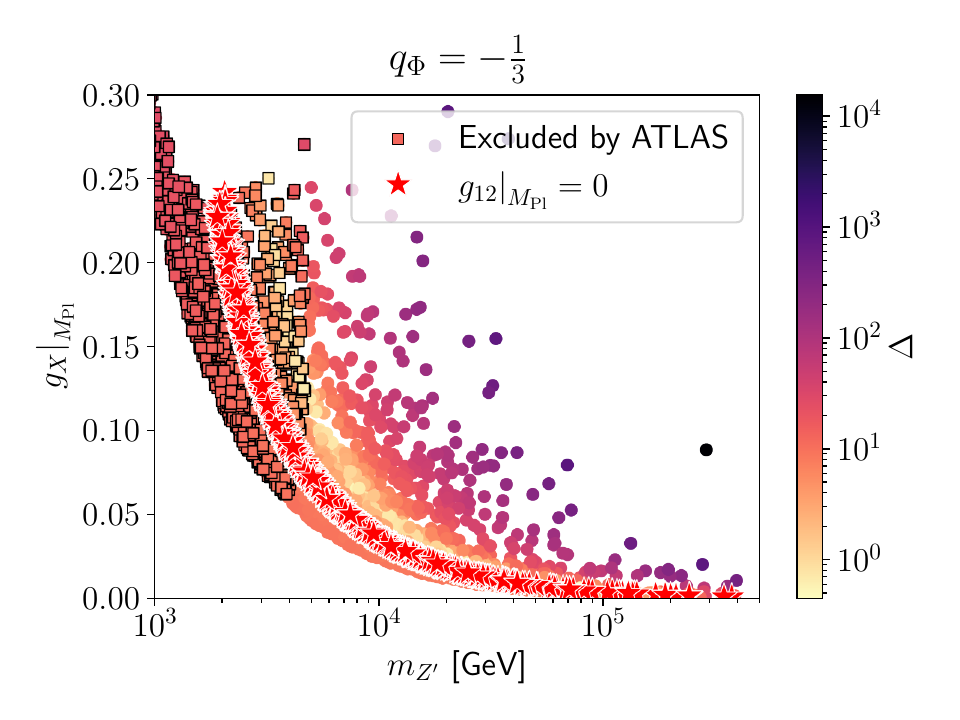}
	\caption{Parameter points that reproduce the correct EW scale in the minimal model. Shown are the custodial symmetry violating parameter $g_{12}$ (left) and the amount of fine tuning (right).}
	\label{fig. minimal model}
\end{figure}
We show the points that reproduce the correct EW scale in Fig.~\ref{fig. minimal model}. Points marked by red stars obey $g_{12}\bigr|_{M_\mathrm{Pl}}=0$ and for these points the new physics scale (say, $m_{Z'}$) is set by $g_X$. Other points have a random value for $g_{12}\bigr|_{M_\mathrm{Pl}}$. The input value of $g_{12}$ plays a major role in the hierarchy between the new physics scale and the EW scale (see Fig.~\ref{fig. minimal model} (left)). For $g_{12}\bigr|_{M_\mathrm{Pl}}\lesssim -0.075\cdot g_X\bigr|_{M_\mathrm{Pl}}$, $\lambda_p$ would be larger than $\lambda_\Phi$, implying that no EWSB would occur which excludes this region. 

\begin{figure}
	\centering
	\includegraphics[width=0.5\textwidth]{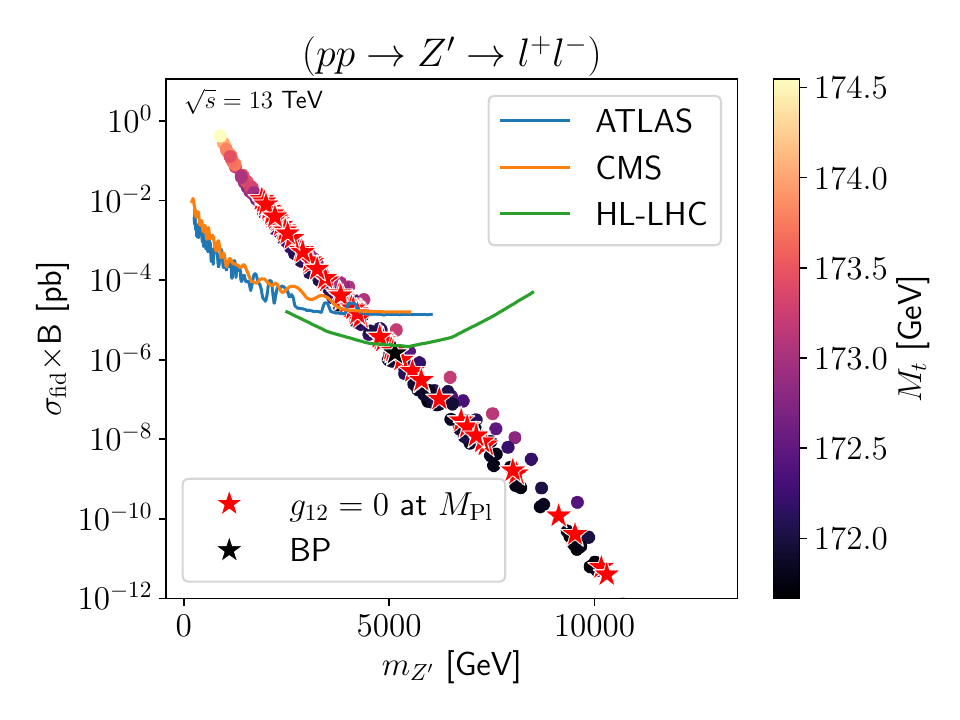}%
	\includegraphics[width=0.5\textwidth]{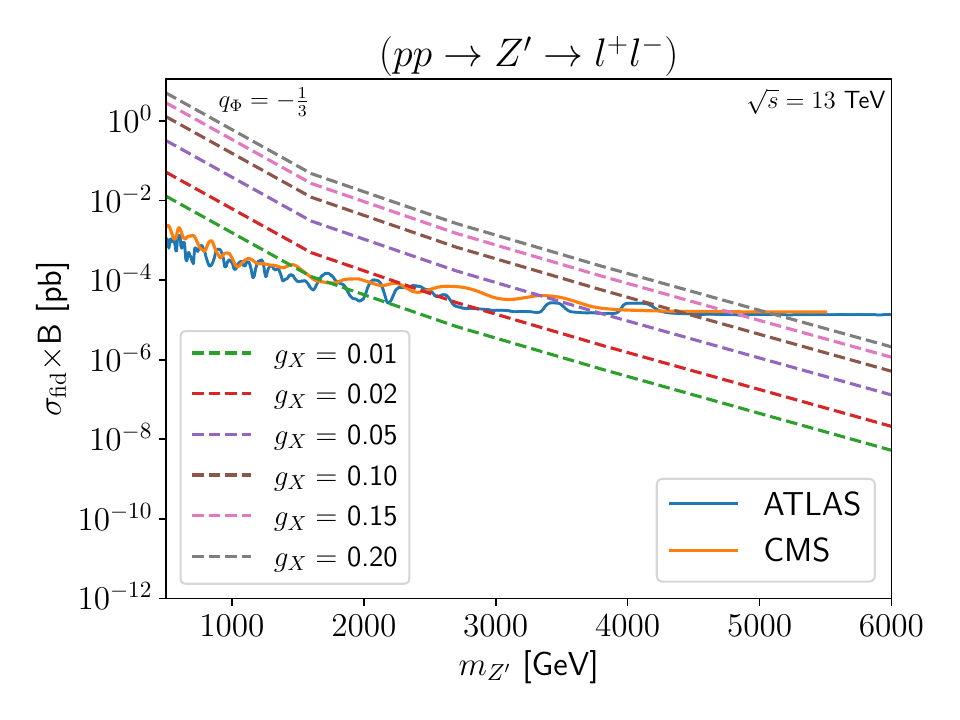}
	\caption{Left: Fiducial cross section times branching ratio ($Z'\rightarrow l^+l^-$ with $l=e,\mu$) for parameter points of the minimal model and $95\%\,\mathrm{C.L.}$ limits from ATLAS and CMS dilepton resonance searches~\cite{ATLAS:2019erb,CMS:2021ctt} as well as projections for HL-LHC (at $14\,\mathrm{TeV}$)~\cite{ATLAS:2018tvr}. Only points that reproduce the correct EW scale and Higgs mass are shown. Right: Fiducial cross section times branching ratio for $Z'$ boson with $q_\Phi=-\frac13$ for different values of $g_X$.}
	\label{fig. fid cross}
\end{figure}

In order to quantify fine tuning, we use a variant of the Barbieri-Giudice measure~\cite{Barbieri:1987fn}. We calculate 
\begin{equation}
 \Delta:=\max_{g_i}\left|\frac{g_i}{\frac{\langle H\rangle}{\langle\Phi\rangle}}\frac{\partial\frac{\langle H\rangle}{\langle\Phi\rangle}}{\partial g_i}\right|=\max_{g_i}\left|\frac{g_i}{\langle H\rangle}\frac{\partial\langle H\rangle}{\partial g_i}-\frac{g_i}{\langle\Phi\rangle}\frac{\partial\langle \Phi\rangle}{\partial g_i}\right|.\label{eq. FT measure}
\end{equation}
The VEVs have a shared sensitivity 
to the high scale, intrinsic to the mechanism of dimensional transmutation.
Our choice of measure in Eq.~\eqref{eq. FT measure} automatically subtracts this common sensitivity, which is not the result of a fine tuning~\cite{Anderson:1994dz}, in order to expose the actual tuning required to obtain the hierarchy between $\langle\Phi\rangle$ and $\langle H\rangle$.
The derivatives in Eq.~(\ref{eq. FT measure}) are calculated numerically by small variations of the input values at $M_\mathrm{Pl}$.
The fine tuning for all points is shown in Fig.~\ref{fig. minimal model} (right). For most points, $\Delta\lesssim 10$ demonstrating that our mechanism generates a hierarchy of $\langle H\rangle\approx 10^{-3}\times \langle \Phi\rangle$ without fine tuning.
The fine tuning measure also has a minimum where $\Delta\lesssim 1$. Whether there is a physical meaning to this ``valley of minimal tuning'' is presently unclear. While the order of magnitude for the tuning is independent of the choice of measure, the valley is not. If we calculate the fine tuning for $m_h/\langle\Phi\rangle$  rather than $\langle H\rangle/\langle\Phi\rangle$, then there is no valley.\\
In order to quantify the strongest direct experimental constraints on this model, we calculate the fiducial cross section times branching ratio for $Z'$ production and decay into two leptons ($l=e,\mu$) using \texttt{MadGraph5\_aMC@NLO}~\cite{Alwall:2014hca} with an \texttt{UFO} file~\cite{Degrande:2011ua} obtained using \texttt{FeynRules}~\cite{Christensen:2008py}. The results, with the same fiducial cuts as in Ref.~\cite{ATLAS:2019erb}, are shown in Fig.~\ref{fig. fid cross} (left). Dilepton resonance searches~\cite{ATLAS:2019erb,CMS:2021ctt} exclude $m_{Z'}\lesssim 4\,\mathrm{TeV}$. We recast the limits from Ref.~\cite{ATLAS:2019erb} by calculating the fiducial cross section times branching ratio for the $Z'$ boson in our model with different masses and couplings (Fig.~\ref{fig. fid cross} (right)). We take the intersections of the lines with fixed $g_X$ and the $95\%\,\mathrm{C.L.}$ exclusion contour of ATLAS~\cite{ATLAS:2019erb}. In case of two crossings, we use the lower value. Interpolating these points then allows us to exclude points on the $m_{Z'}-g_X$ plane. The excluded points are marked by black squares in Fig.~\ref{fig. minimal model} and the rest of this work. While Fig.~\ref{fig. fid cross} shows the results for $q_\Phi=-\frac13$, we also do the same calculation with $q_\Phi=-\frac38$ to obtain the recasted limits for models with $q_\Phi=-\frac38$.\footnote{We do not include new fermions in the calculation. These new fermions affect the width of the $Z'$ boson, however the effect is small.}

\subsection{Minimal fermion extension - Neutrino portal model}\label{sec. neutrino model}
Next, we consider a model with the choice $q_\Phi=-\frac38$. For this charge assignment, gauge kinetic mixing tends to remain small (see Fig.~\ref{fig. stream plot}). The minimal setup with $g_{12}=0$ at the Planck scale does not lead to EWSB since the gauge kinetic mixing remains too small to overcome the SM contributions to $\beta_{\lambda_p}-\beta_{\lambda_\Phi}$ which have the incorrect sign. Therefore, additional sources of custodial symmetry violation are required. The simplest way is to assume $g_{12}\bigr|_{M_\mathrm{Pl}}>0$. Alternatively, we can introduce new fermions with a Yukawa coupling to $\Phi$.  

The minimal way to introduce new fermions while allowing for a Yukawa interaction involving $\Phi$ is shown in Tab.~\ref{tab. particle content U1 model}~(middle). The new fermions are vector-like\footnote{All vector-like mass term are forbidden by scale invariance.} and their contributions to gauge anomalies cancel. 
The new Yukawa interaction given in Eq.~(\ref{eq. Yukawa psi nuR}) involves $\psi_L$, $\Phi$ and $\nu_R$, connecting the new physics sector to the neutrino portal.

After spontaneous symmetry breaking, the neutral fermions obtain Dirac mass terms given by
\begin{equation}
	\mathcal{L}_\mathrm{mass}\supset\left(\begin{matrix}\overline\nu_L^\alpha & \overline\psi_L\end{matrix}\right)\left(\begin{matrix}y_\nu^{\alpha\beta} \frac{v_H}{\sqrt{2}} & 0 \\y_\psi^\beta \frac{v_\Phi}{\sqrt{2}} & 0\end{matrix}\right)\left(\begin{matrix}\nu_R^\beta \\ \psi_R\end{matrix}\right)+\text{h.c.}~=:~\left(\begin{matrix}\overline\nu_L^\alpha & \overline\psi_L\end{matrix}\right)M_N\left(\begin{matrix}\nu_R^\beta \\ \psi_R\end{matrix}\right)+\text{h.c.} \label{eq. nu mass mat}
\end{equation}
Majorana mass terms are not generated even at loop level due to an unbroken (accidental) lepton number symmetry.
An additional chiral symmetry ensures that $\psi_R$ cannot obtain a mass term. The squared fermion masses are obtained as the eigenvalues of ($\alpha,\alpha'=1,2,3$, a sum over $\beta$ is implicit)
\begin{equation}
	M_N M_N^\dagger=\left(\begin{matrix}y_\nu^{\alpha\beta}(y_\nu^\dagger)^{\beta\alpha'}\frac{v_H^2}{2} & y_\nu^{\alpha\beta}(y^*_\psi)^{\beta}\frac{v_Hv_\Phi}{2}\\ y_\psi^{\beta}(y^\dagger_\nu)^{\beta\alpha'}\frac{v_Hv_\Phi}{2}& y_\psi^\beta (y^*_\psi)^\beta\frac{v_\Phi^2}{2}\end{matrix}\right).
\end{equation}
For simplicity, we assume real Yukawa couplings. The mass matrix has rank 3 (see Eq.~\ref{eq. nu mass mat}), i.e.\ one eigenvalue vanishes. Since $v_\Phi\gg v_H$, the lower right entry dominates and the heavy sterile (with respect to the SM interactions) eigenstate with mass $\approx \sqrt{y_\psi^\beta y_\psi^\beta}\frac{v_\Phi}{\sqrt{2}}=\bar y_\psi\frac{v_\Phi}{\sqrt{2}}$, written as a Dirac spinor $\Psi$, is given by
\begin{equation}
	\Psi\sim\left(\begin{matrix}\cos(\alpha_\psi)\psi_L+\sin(\alpha_\psi)\nu_L\\ \nu'_R\end{matrix}\right).
\end{equation}
Here, $\sin(\alpha_\psi)\approx y_\nu v_H/(y_\psi v_\Phi)$, which is automatically suppressed thereby justifying the notion as a sterile state, and $\nu'_R$ is a linear combination of the right-handed neutrinos (not involving $\psi_R$). 
The other two massive eigenstates are active Dirac neutrinos with masses $\sim y_\nu\frac{v_H}{\sqrt{2}}$. The remaining massless state is an active neutrino, i.e.\ this model predicts that the lightest generation of active neutrinos is massless.

\begin{figure}
	\centering
	\includegraphics[width=0.5\textwidth]{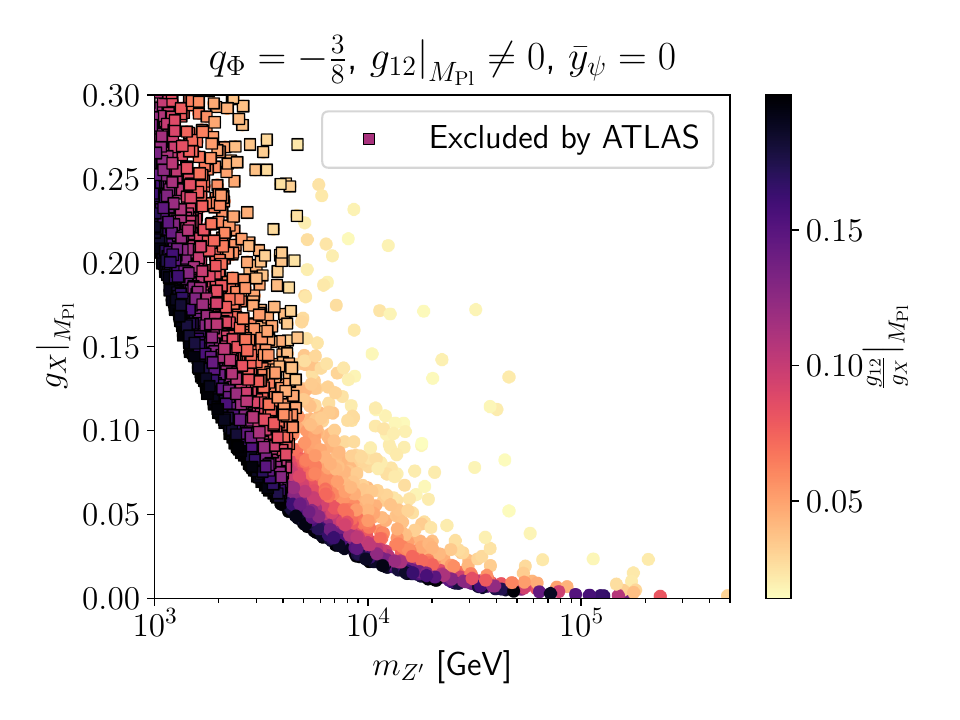}%
	\includegraphics[width=0.5\textwidth]{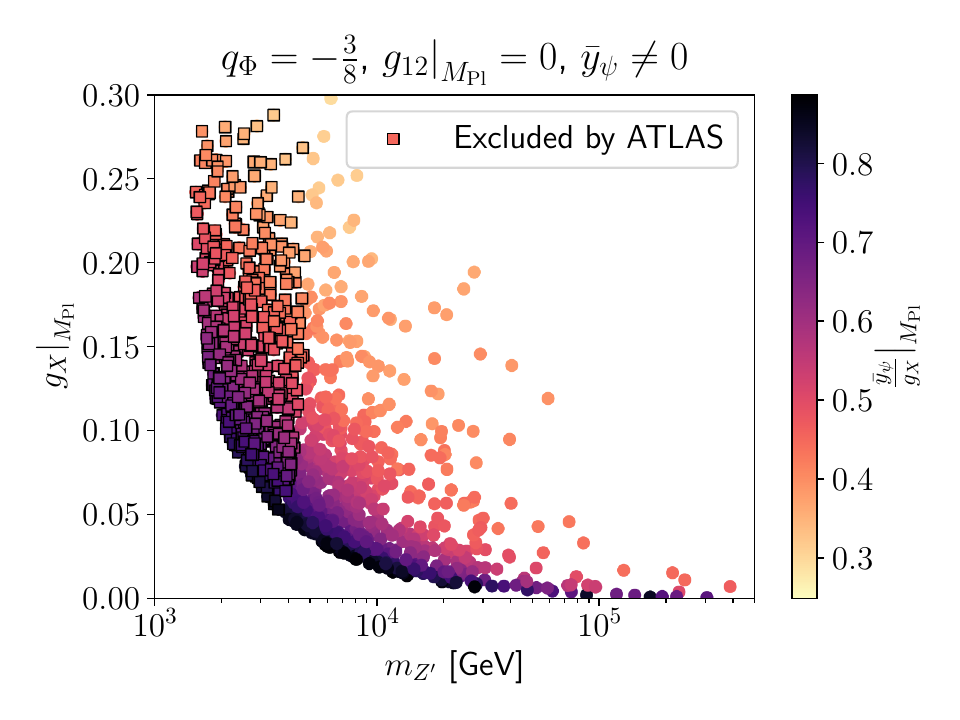}
	\includegraphics[width=0.5\textwidth]{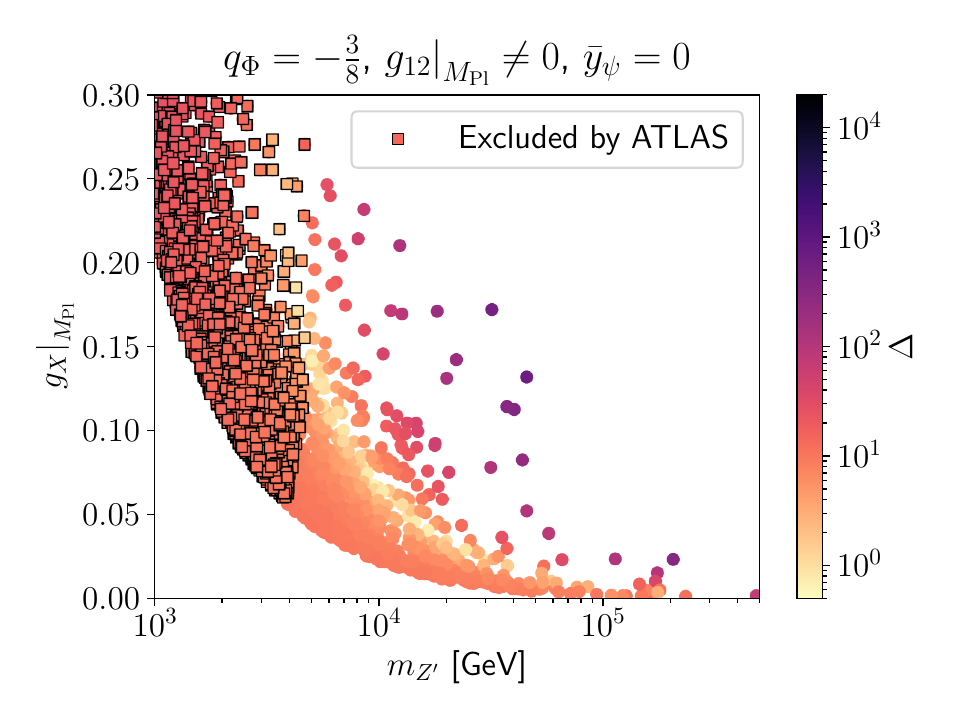}%
	\includegraphics[width=0.5\textwidth]{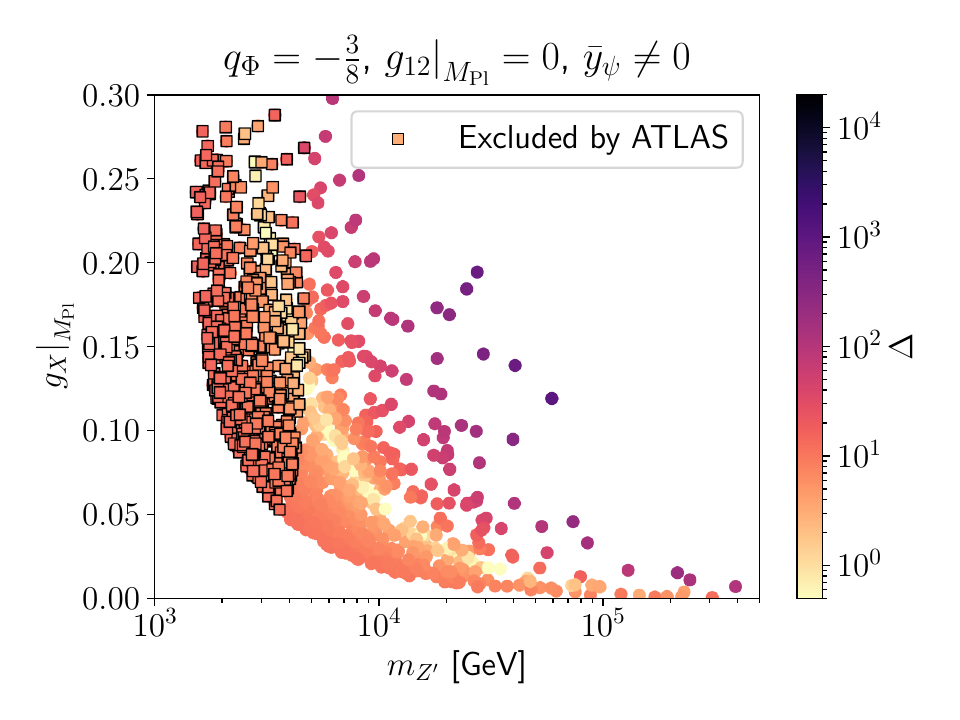}
	\caption{Parameter points that reproduce the correct EW scale in the neutrino portal model with $\bar y_\psi^\alpha=0$ and $g_{12}\bigr|_{M_\mathrm{Pl}}\neq 0$ (left) and with $\bar y_\psi^\alpha\neq0$, $g_{12}\bigr|_{M_\mathrm{Pl}}= 0$ (right). Shown are the effects of the custodial symmetry violation via $g_{12}$ and $\bar y_\psi$ (top) and the amount of fine tuning (bottom).}
	\label{fig. neutrino model}
\end{figure}

We perform two parameter scans with a setup similar to the minimal model. For the first scan, we assume $\bar y_\psi=0$ and $g_{12}\bigr|_{M_\mathrm{Pl}}\in[0,0.2]\cdot g_X\bigr|_{M_\mathrm{Pl}}$ and for the second scan  $\bar y_\psi\bigr|_{\tilde\mu_0}\in[0,0.9]\cdot g_X\bigr|_{\tilde\mu_0}$ and $g_{12}\bigr|_{M_\mathrm{Pl}}=0$. Fig.~\ref{fig. neutrino model} shows how the strength of custodial symmetry violation affects the hierarchy, illustrating how small but non-zero values of $\bar y_\psi$ or $g_{12}\bigr|_{M_\mathrm{Pl}}$ are required to overcome the SM contributions to $\lambda_p-\lambda_\Phi$. Large values of $\bar y_\psi$ or $g_{12}\bigr|_{M_\mathrm{Pl}}$ lead to more \SO{6} custodial symmetry violation and therefore a smaller hierarchy. This is a numerical demonstration of the fact that additional sources of custodial symmetry breaking are needed for successful phenomenology.
We also show the amount of fine tuning calculated using Eq.~(\ref{eq. FT measure}).

\subsection{Dark matter model}
\begin{figure}
	\centering
    \includegraphics[width=0.45\textwidth]{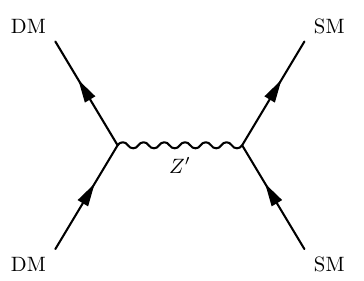}%
    \hspace{1cm}%
    \includegraphics[width=0.45\textwidth]{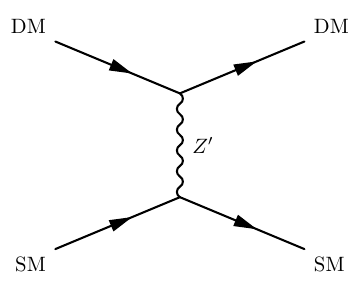}
	\caption{Left: Feynman diagram for dominant DM annihilation channel. The final states are SM fermions. Right: Feynman diagram for the dominant channel contributing to DM-nucleon scattering.}
	\label{fig. DM diagrams}
\end{figure}

We present a simple model that allows for two-component WIMP DM\footnote{The minimal way to include new stable fermions while canceling the gauge anomalies requires us to introduce two copies. These stable particles form two-component DM meaning that both components will have a non-zero relic density.} for which we again choose $q_\Phi=-\frac38$. Similar to the model in Sec.~\ref{sec. neutrino model}, gauge kinetic mixing remains small (see Fig.~\ref{fig. stream plot}). We add a pair of vector-like fermions with charges given in Tab.~\ref{tab. particle content U1 model} (bottom). The value of $p$ is a free parameter as long as the coupling to the right-handed neutrinos is forbidden. For the numerical analysis we choose $p=\frac12$.
The new Yukawa interactions are given in Eq.~(\ref{eq. Yukawa psi psip}) and both new Yukawa interactions contribute to custodial symmetry violation. The Lagrangian has two global (accidental) \U{1} symmetries under which only the new fermions are charged. 
Since both of these symmetries remain unbroken after the scalar fields obtain their VEVs, both $\psi$ and $\psi'$ are stable and make up two-component DM. 
In the early Universe, $\psi$ and $\psi'$ are in thermal equilibrium with the SM and the DM relic density is obtained via a freeze-out process, separately for each of the DM components.\footnote{Thermal equilibrium is only reached if the reheating temperature is high enough which is not necessarily the case in scale invariant models (see e.g.\ Ref.~\cite{Konstandin:2011dr} and our discussion in Sec.~\ref{sec. finite T}). A proper analysis of the DM phenomenology requires a detailed analysis of the thermal history of the Universe.} The total DM relic density $\Omega h^2$ is then given by the sum of the individual relic densities, i.e $\Omega h^2=(\Omega h^2)_\psi+(\Omega h^2)_{\psi'}$ . The dominant annihilation diagram is the s-channel $Z'$ exchange with SM fermions in the final state (see Fig.~\ref{fig. DM diagrams}). 

\begin{figure}
	\centering
	\includegraphics[width=0.5\textwidth]{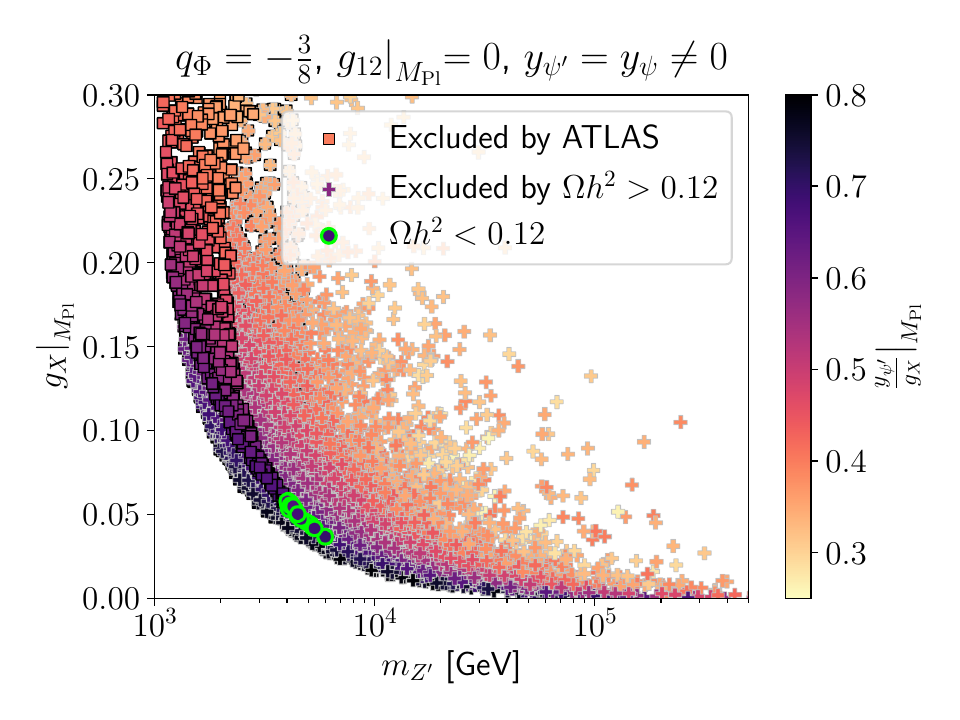}%
	\includegraphics[width=0.5\textwidth]{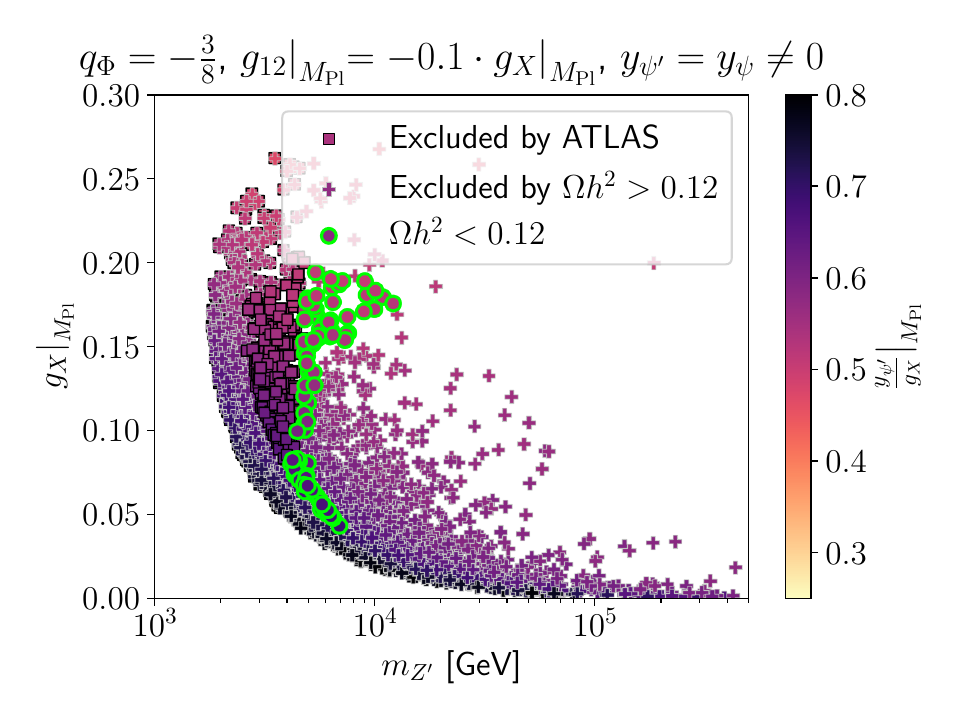}
	\includegraphics[width=0.5\textwidth]{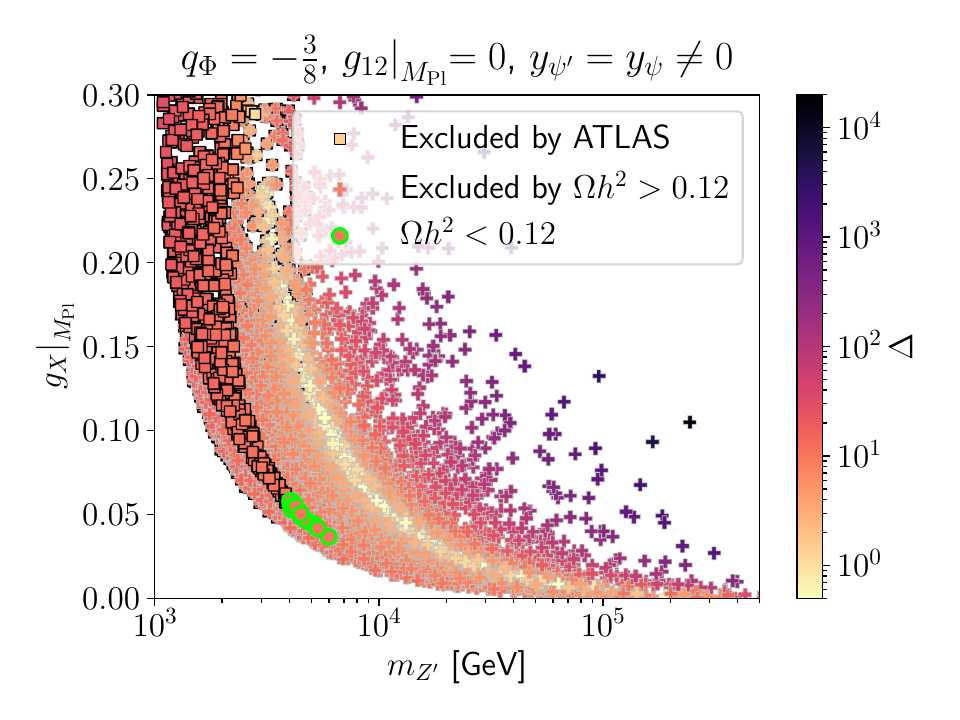}%
	\includegraphics[width=0.5\textwidth]{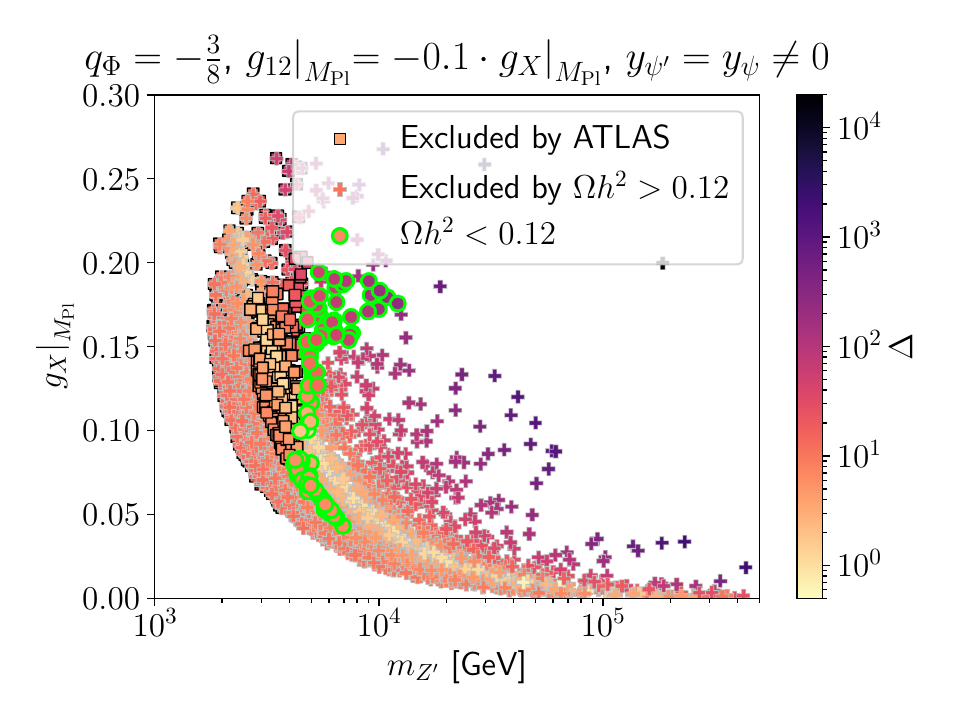}
	\includegraphics[width=0.5\textwidth]{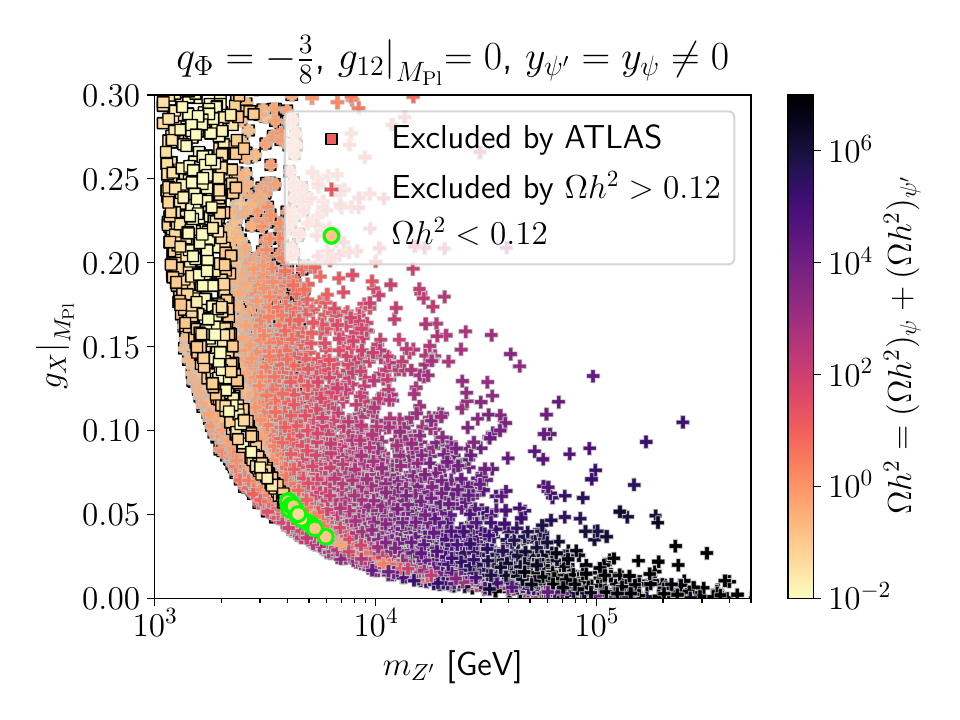}%
	\includegraphics[width=0.5\textwidth]{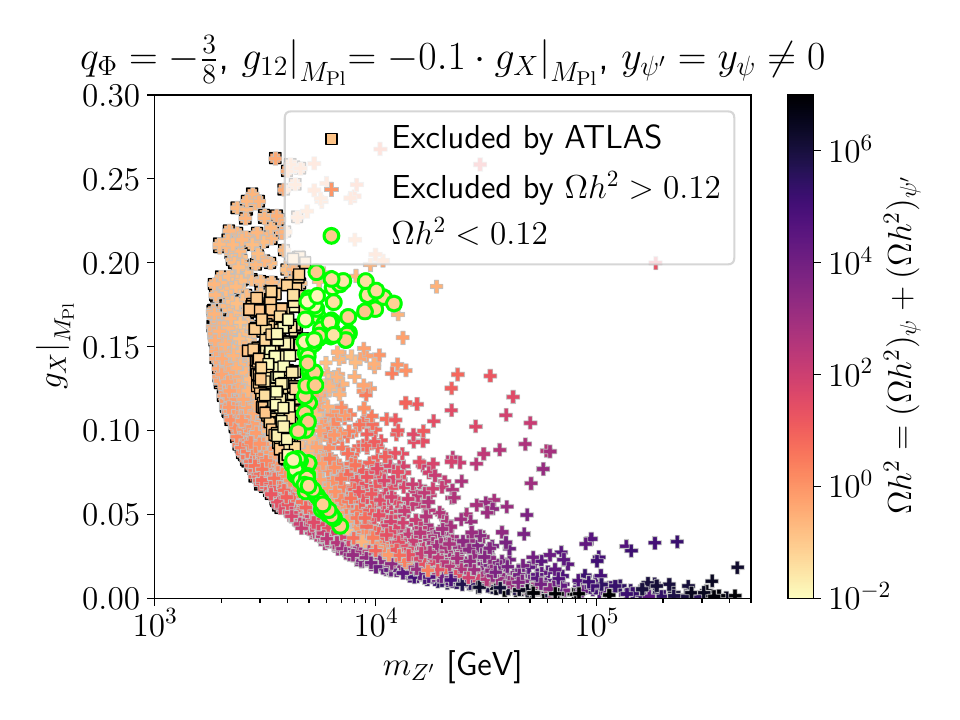}
	\caption{Parameter points for the DM model which reproduce the correct EW scale assuming $g_{12}\bigr|_{M_\mathrm{Pl}}=0$ (left) and $g_{12}\bigr|_{M_\mathrm{Pl}}=-0.1\cdot g_X\bigr|_{M_\mathrm{Pl}}$ (right). The two Yukawa couplings are, for simplicity, chosen as $y_{\psi'}= y_\psi$. Shown are the effects of custodial symmetry violation via $y_\psi$ and $y_{\psi'}$ (top), the fine tuning (middle) and the DM relic density (bottom). Points that yield a DM relic density above the observed value of $\Omega h^2=0.12$~\cite{Planck:2018vyg} are excluded and marked as gray bordered plus symbols while points that yield $\Omega h^2<0.12$, and are not excluded by the ATLAS dilepton resonance searches, are marked with bright green bordered points.}
	\label{fig. DM plots}
\end{figure}

We use the same numerical setup as in the previous sections. We randomly choose $y_\psi\bigr|_{\tilde\mu_0}\in[0,0.8]\cdot g_X\bigr|_{\tilde\mu_0}$ and for simplicity $y_{\psi'}= y_\psi$. We perform two parameter scans, one with $g_{12}\bigr|_{M_\mathrm{Pl}}=0$ and the second one uses $g_{12}\bigr|_{M_\mathrm{Pl}}=-0.1\cdot g_X\bigr|_{M_\mathrm{Pl}}$. In the second case, $g_{12}$ and $y_\psi,y_{\psi'}$ have opposite contributions to $\lambda_p-\lambda_\Phi$, thereby allowing for larger values of $m_\psi/m_{Z'}$. The effect of custodial symmetry violation from $y_\psi=y_{\psi'}$ and the fine tuning is shown in Fig.~\ref{fig. DM plots} (top and middle). The fine tuning for the parameter points with $g_{12}\bigr|_{M_\mathrm{Pl}}=-0.1\cdot g_X\bigr|_{M_\mathrm{Pl}}$ is slightly larger than for $g_{12}\bigr|_{M_\mathrm{Pl}}=0$ since gauge kinetic mixing partially cancels the effects of $y_\psi$ and $y_\psi'$ and this cancellation requires tuning.
We calculate the DM relic density using \texttt{micrOMEGAs 6.0.5}~\cite{Alguero:2023zol} with model files generated using \texttt{SARAH-4.15.2}~\cite{Staub:2013tta}. The results for the relic density are shown in Fig.~\ref{fig. DM plots} (bottom). For large parts of the parameter space, the relic density is too large. However, near the $Z'$ resonance $m_\psi\approx m_{\psi'}\approx\frac12 m_{Z'}$, the annihilation rate is sufficiently large and the relic density can reach the observed or smaller values. Note that $y_\psi\approx y_{\psi'}$, up to deviations of a few percent, is required so that both DM candidates can be near the $Z'$ resonance simultaneously.\footnote{$y_\psi=y_{\psi'}$ can be justified by a parity-type symmetry that maps $\psi_L\leftrightarrow \psi'_R$ and $\psi'_L\leftrightarrow \psi_R$.} For $g_{12}\bigr|_{M_\mathrm{Pl}}=0$ most of the points that do not overproduce DM are excluded be the ATLAS searches. Only a very small region in parameter space is not excluded by these constraints. Negative values for $g_{12}\bigr|_{M_\mathrm{Pl}}$ allow for higher $Z'$ masses while not overproducing DM. Points that yield the correct relic abundance lie on the boundary of the region with $\Omega h^2<0.12$. 
\begin{figure}
	\centering
	\includegraphics[width=0.5\textwidth]{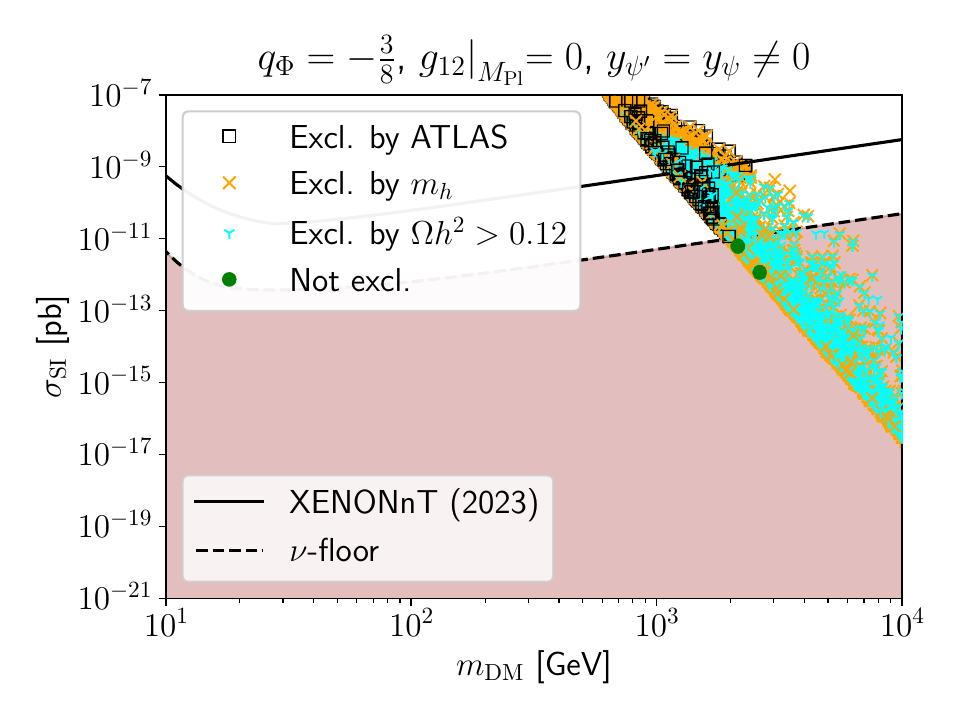}%
	\includegraphics[width=0.5\textwidth]{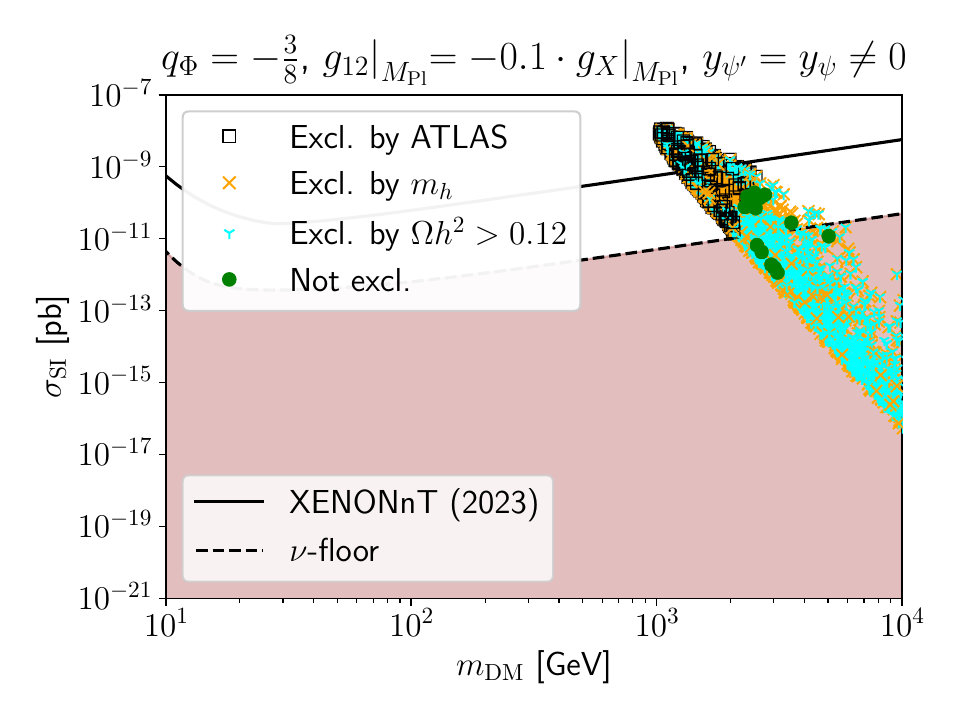}
	\caption{
	Spin independent (SI) cross section for WIMP-nucleon scattering and XENONnT limits~\cite{XENON:2023cxc}. Also shown is the neutrino floor for Xenon~\cite{OHare:2021utq}. Black squares indicate points excluded by the ATLAS dilepton resonance searches, orange crosses indicate points where the Higgs mass is outside of its $3\sigma$ range, and cyan ``tri-down'' points indicate points with a DM relic abundance $\Omega h^2>0.12$. Phenomenologically viable points are labeled by green dots.}
	\label{fig. SI cross}
\end{figure}	
Our DM candidates can scatter off nuclei via a virtual $Z'$ exchange (see Fig.~\ref{fig. DM diagrams}).
We calculate the direct detection cross section using \texttt{micrOMEGAs 6.0.5}. Both, $\psi$ and $\psi'$ have the same cross section (assuming $m_\psi=m_{\psi'}$) and the results for the spin independent scattering cross section $\sigma_\mathrm{SI}=\sigma_\mathrm{SI}(\psi)=\sigma_\mathrm{SI}(\psi')$ are shown in Fig.~\ref{fig. SI cross}. Current direct detection limits turn out to be weaker than the ATLAS limits on the $Z'$ boson. Future direct detection experiments such as DARWIN will reach the neutrino floor and probe parts of our parameter space~\cite{DARWIN:2016hyl}.
More detailed analyses of fermionic DM models with a massive $Z'$ portal can be found e.g.\ in Refs.~\cite{Alves:2013tqa,Alves:2015pea,Wang:2015saa,Alves:2015mua,Jacques:2016dqz,Okada:2018ktp}.

\section{New particle masses and further experimental signatures}\label{sec. constraints and signatures}
\begin{figure}
	\centering
	\includegraphics[width=0.5\textwidth]{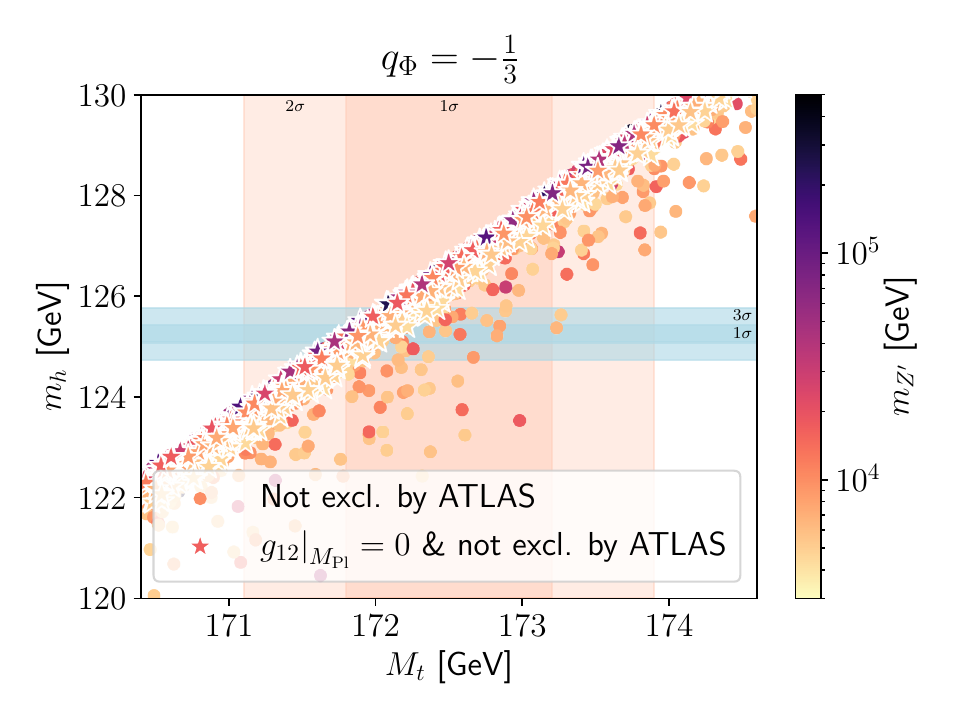}%
	\includegraphics[width=0.5\textwidth]{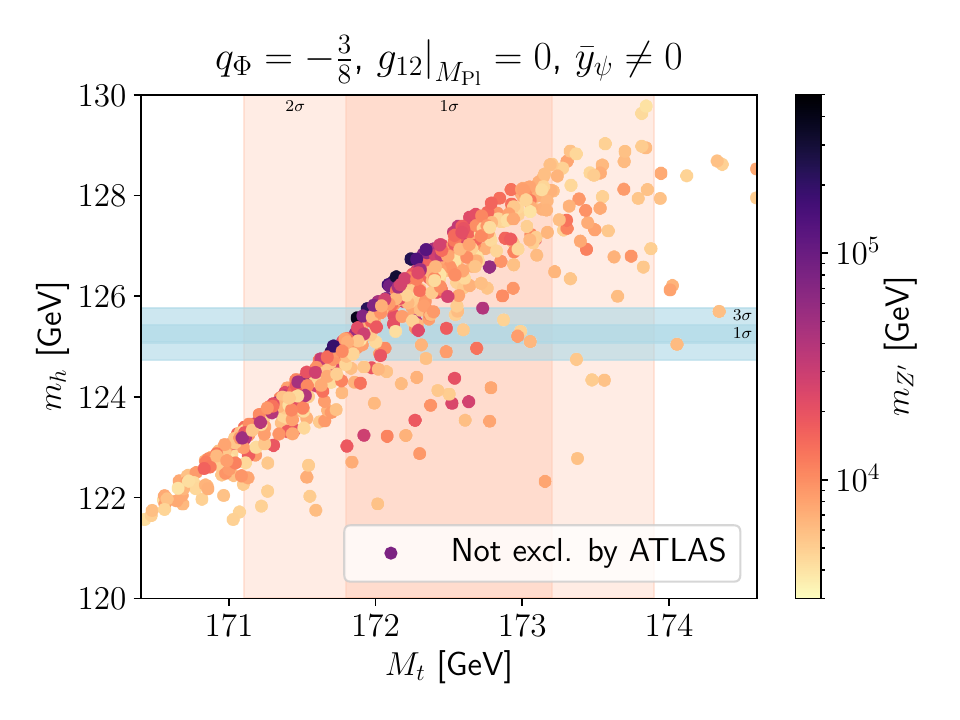}
	\includegraphics[width=0.33\textwidth]{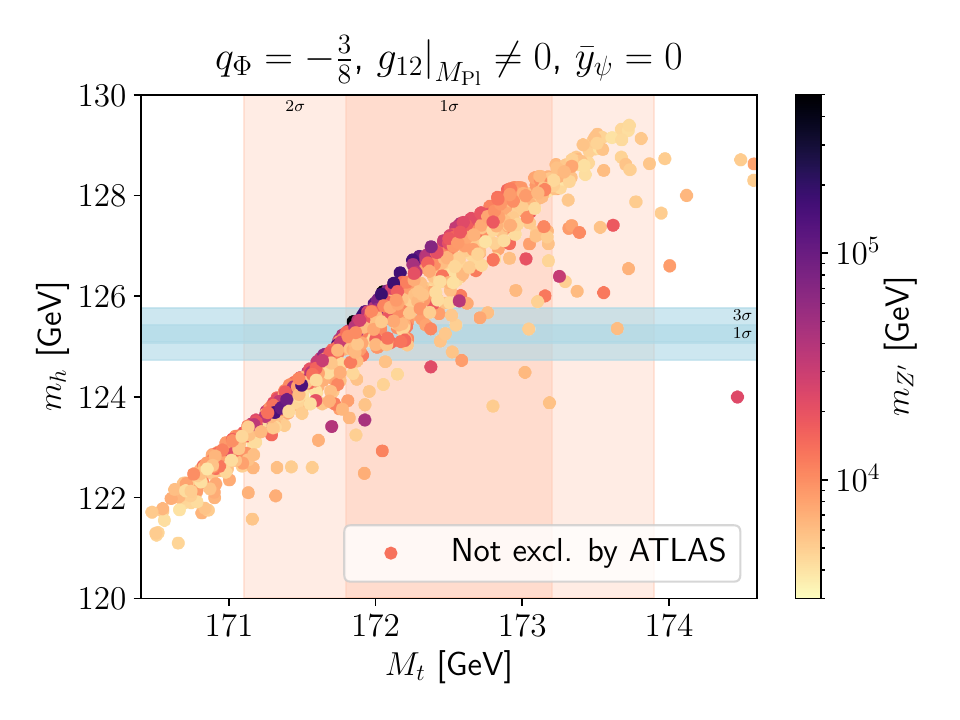}%
	\includegraphics[width=0.33\textwidth]{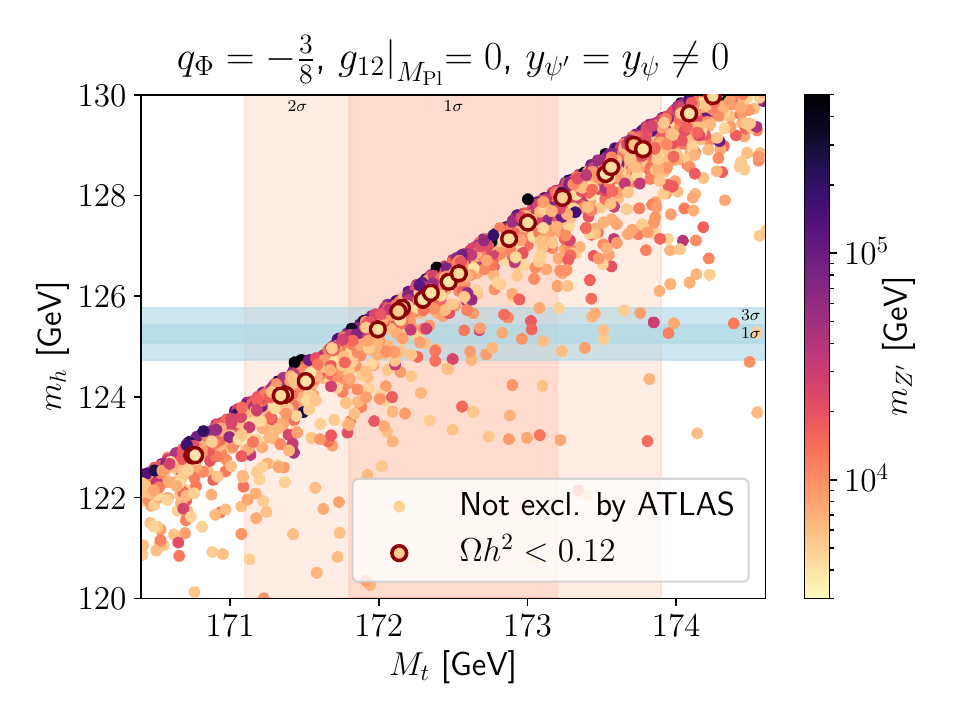}%
	\includegraphics[width=0.33\textwidth]{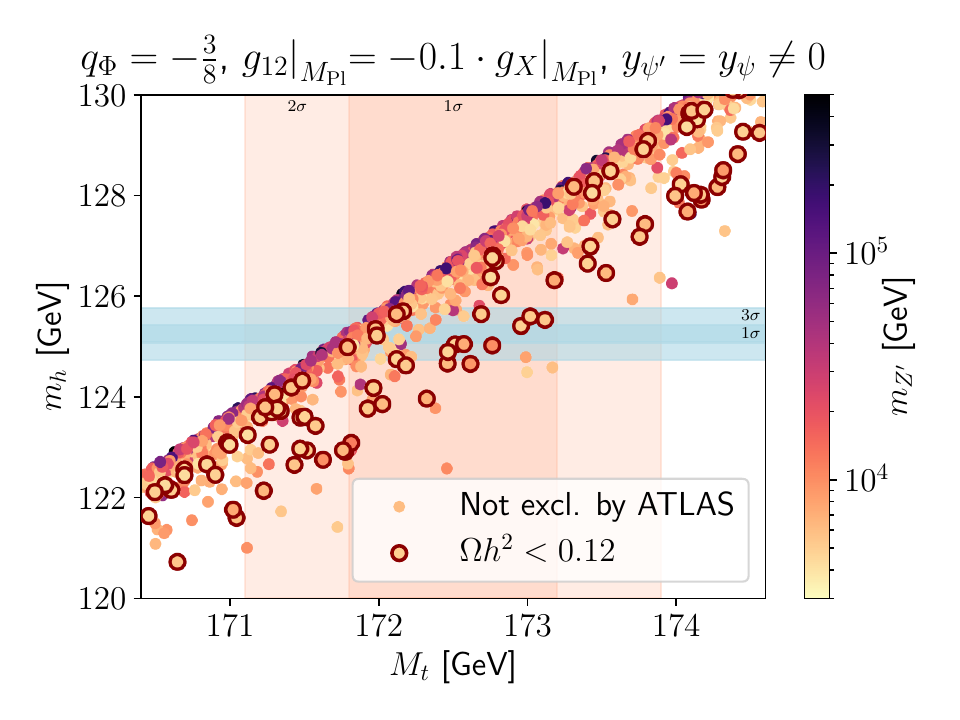}%
	\caption{Correlation of the top pole mass $M_t$ and the Higgs mass $m_h$ for the minimal model (top left), the neutrino portal model with $y_\psi\neq0$, $g_{12}\bigr|_{M_\mathrm{Pl}}=0$ (top right) and with $y_\psi\neq0$, $g_{12}\bigr|_{M_\mathrm{Pl}}=0$ (bottom left) and for the DM model with $g_{12}\bigr|_{M_\mathrm{Pl}}=0$ (bottom middle) and with $g_{12}\bigr|_{M_\mathrm{Pl}}=-0.1\cdot g_X\bigr|_{M_\mathrm{Pl}}$ (bottom right) for points that are not excluded by current ATLAS dilepton resonance searches. For the DM model, points that fulfill $\Omega h^2<0.12$ are indicated by dark bordered points.}
	\label{fig. Mtmh}
\end{figure}

The experimental uncertainty of the top quark mass 
turns out to be the major limitation for more precise predictions 
in our model. 
This is because the top Yukawa coupling gives the leading contribution to $\beta_{\lambda_H}$ and the uncertainty, amplified by the running over many orders of magnitude, translates to an uncertainty on the theoretical prediction of the Higgs mass. We show the relation between the numerical values of the top pole mass $M_t$ and the Higgs mass in Fig.~\ref{fig. Mtmh} for all points not excluded by the ATLAS dilepton resonance searches.
 The results are similar for all models studied in this work. The majority of points with the correct Higgs mass also require the top mass to be in its $1\sigma$ range with only very few points reaching the upper end of its $3\sigma$ allowed interval. In the minimal setup with $q_\Phi=-\frac13$ and $g_{12}\bigr|_{M_\mathrm{Pl}}=0$ (white bordered stars), there is an approximately linear relation between the top mass and the Higgs mass. While this setup has the same number of free parameters as the SM, one would need to measure the top mass with $0.1\,\mathrm{GeV}$ precision in order to accurately predict the value of $m_{Z'}$. With higher precision on $M_t$, it might become necessary to calculate the effective potential as well as the RGEs at higher loop order. In the DM model, only few points obey $\Omega h^2<0.12$ (dark bordered points) and these few points seem to have the same distribution as the points with no constraint on the relic density. For all models, there are no viable points with $M_t\lesssim 171.5\,\mathrm{GeV}$. In general, measuring the top quark mass more precisely is an important check for Custodial Naturalness. 

\begin{figure}
	\centering
	\includegraphics[width=0.5\textwidth]{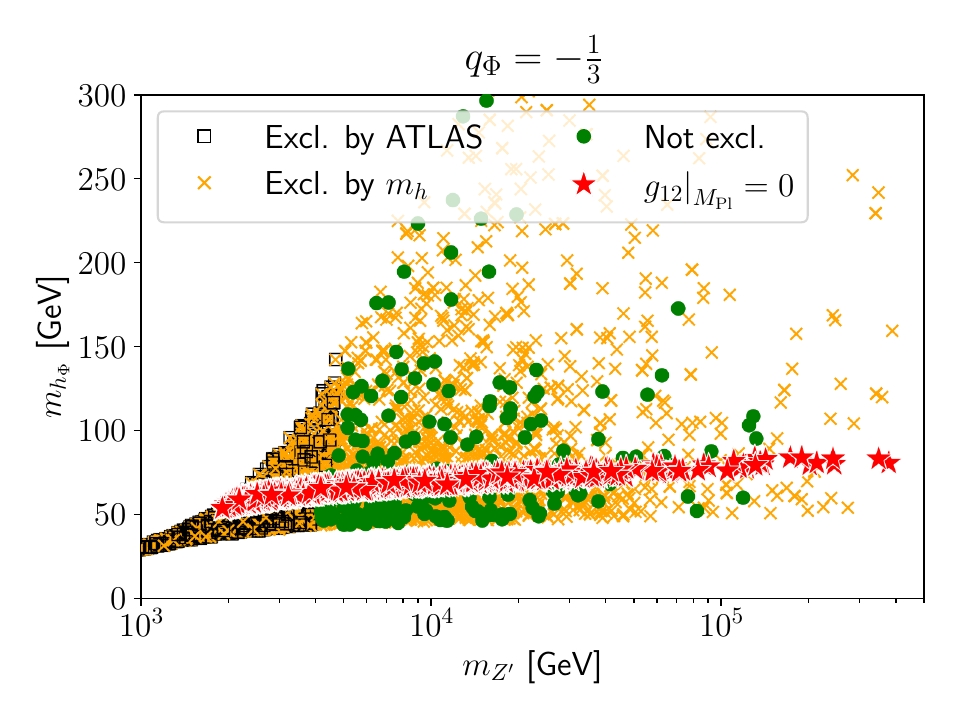}%
	\includegraphics[width=0.5\textwidth]{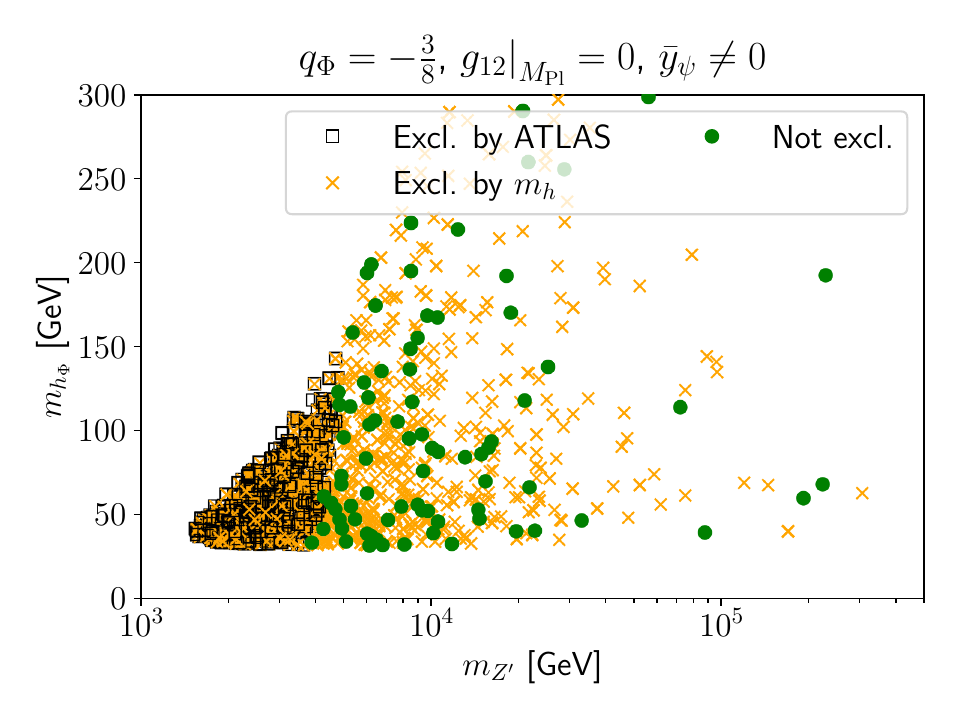}
	\includegraphics[width=0.33\textwidth]{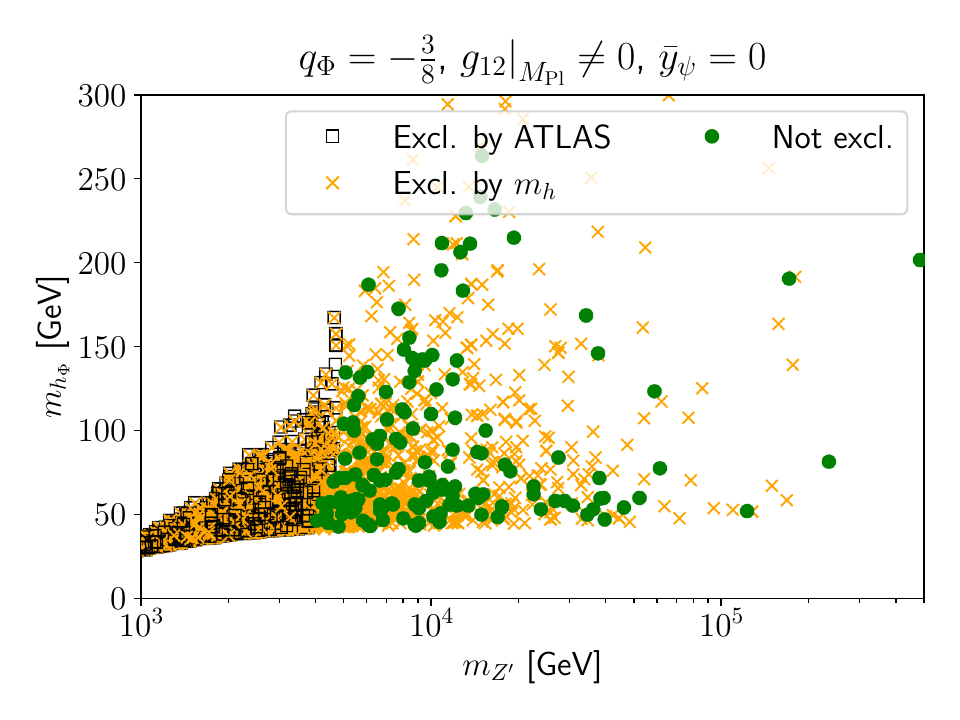}%
	\includegraphics[width=0.33\textwidth]{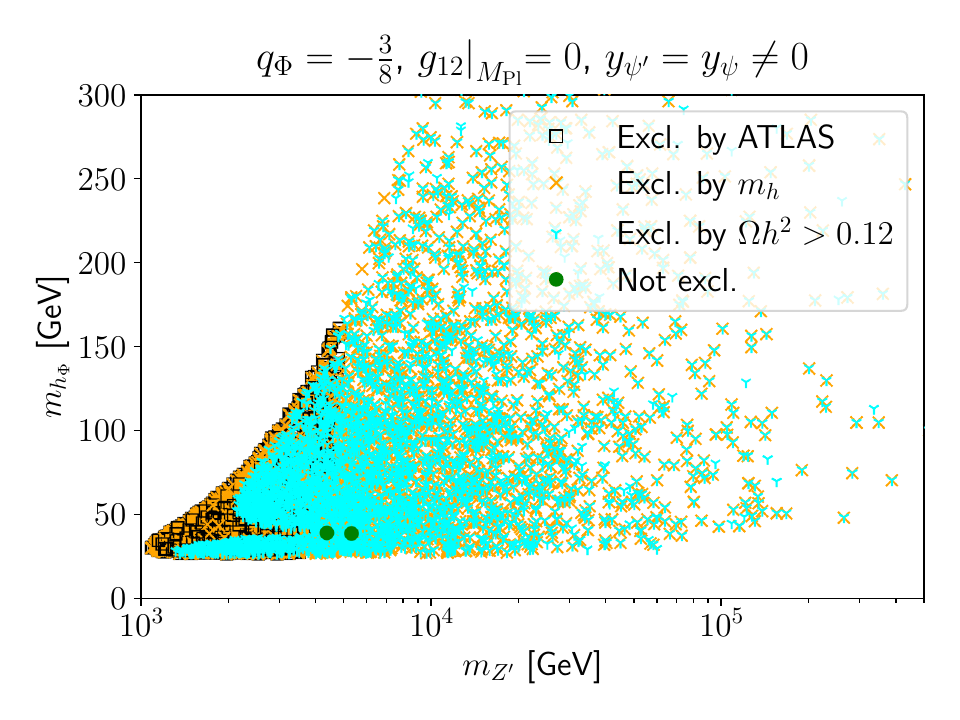}%
	\includegraphics[width=0.33\textwidth]{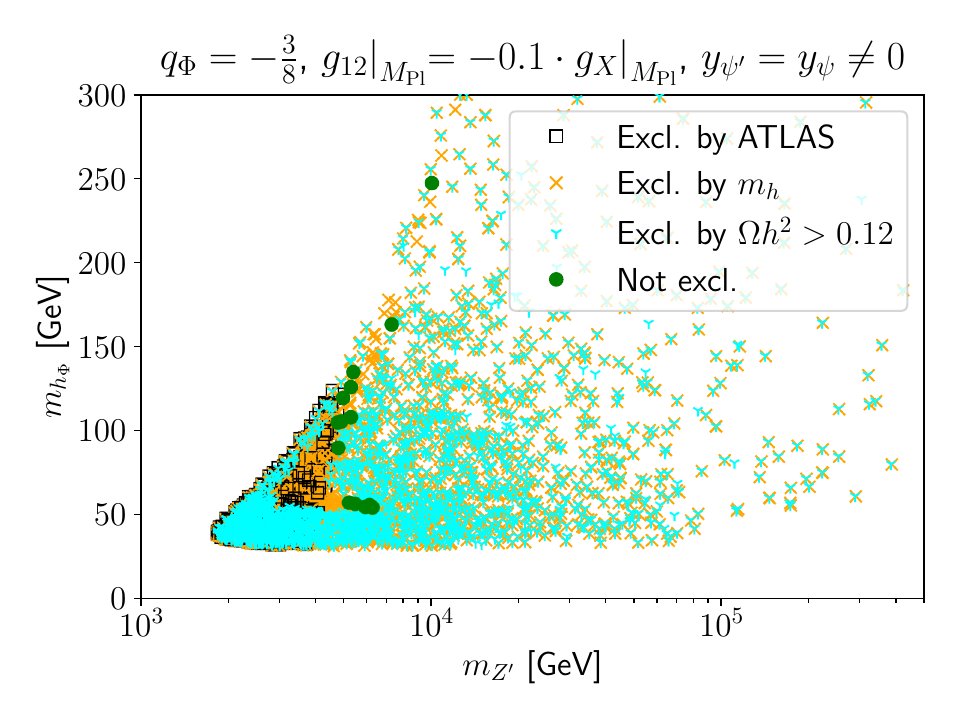}%
	\caption{Numerical values of the dilaton  mass $m_{h_\Phi}$ for the minimal model (top left), the neutrino portal model with $y_\psi\neq0$, $g_{12}\bigr|_{M_\mathrm{Pl}}=0$ (top right) and with $y_\psi\neq0$, $g_{12}\bigr|_{M_\mathrm{Pl}}=0$ (bottom left) and for the DM model with $g_{12}\bigr|_{M_\mathrm{Pl}}=0$ (bottom middle) and with $g_{12}\bigr|_{M_\mathrm{Pl}}=-0.1\cdot g_X\bigr|_{M_\mathrm{Pl}}$ (bottom right). The black squares indicate points excluded by the ATLAS dilepton resonance searches and the orange crosses indicate points where the Higgs mass is outside of its $3\sigma$ range. The green points are not excluded. For the DM model, the cyan ``tri-down'' points indicate the region where the relic abundance is $\Omega h^2>0.12$.}
	\label{fig. mhPhi}
\end{figure}

The particle spectrum of Custodial Naturalness includes the dilaton with a mass suppressed by the beta-function $\beta_{\lambda_\Phi}$ (see Eq.~(\ref{eq. dilaton mass})). We show the values for $m_{h_\Phi}$ in Fig.~\ref{fig. mhPhi}.
 In the minimal scenario with $q_\Phi=-\frac13$ and $g_{12}\bigr|_{M_\mathrm{Pl}}=0$ (red stars), the dilaton mass is always smaller than the Higgs mass and approximately $70\,\mathrm{GeV}$ with only a small dependence on the intermediate scale $\langle\Phi\rangle$. For models with vanishing $y_\psi$ the dilaton mass is bounded from below by $m_{h_\Phi}\gtrsim 40\,\mathrm{GeV}$, while for $y_\psi\neq0$ smaller values are possible only limited by the numerical range of $y_\psi$. In either case, the dilaton mass can reach up to a few $100\,\mathrm{GeV}$. Points that allow for the correct Higgs mass are evenly distributed.

\begin{figure}
	\centering
	\includegraphics[width=0.5\textwidth]{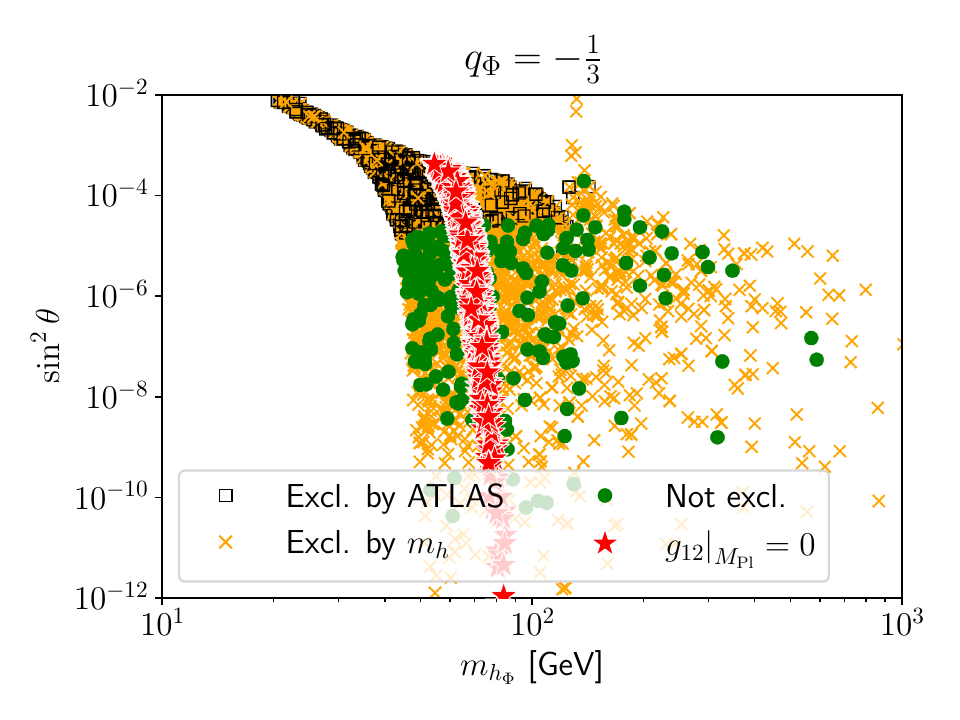}%
	\includegraphics[width=0.5\textwidth]{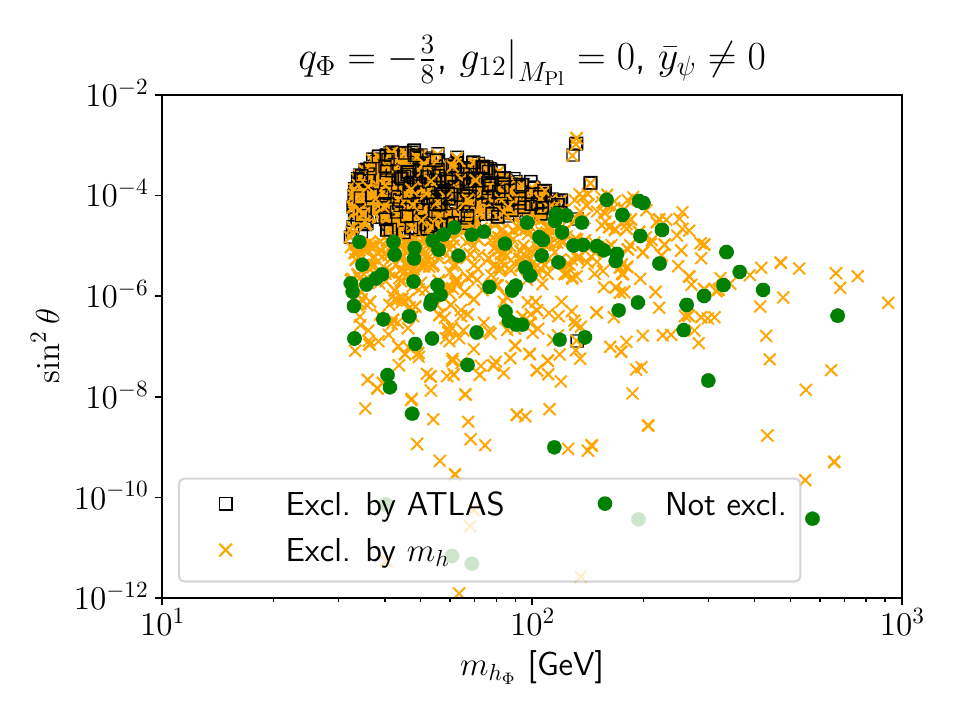}
	\includegraphics[width=0.33\textwidth]{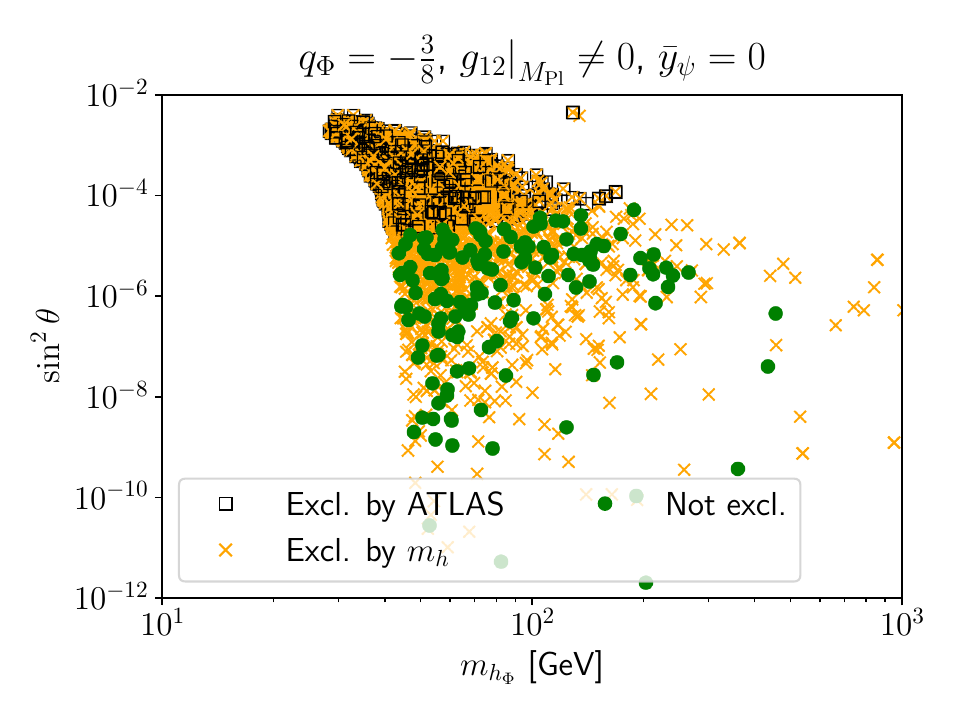}%
	\includegraphics[width=0.33\textwidth]{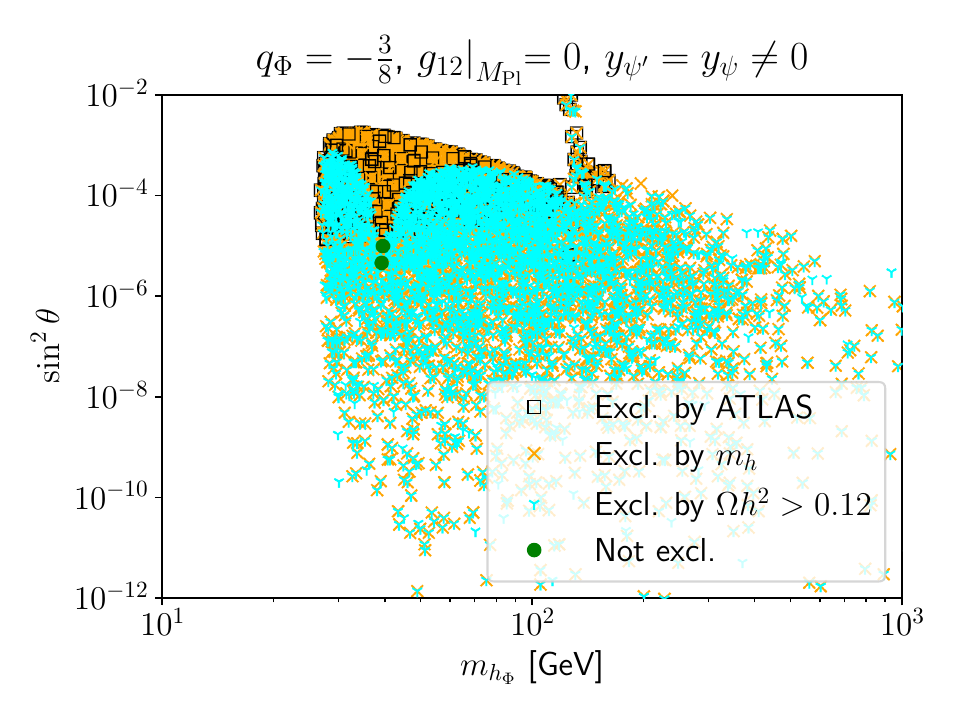}%
	\includegraphics[width=0.33\textwidth]{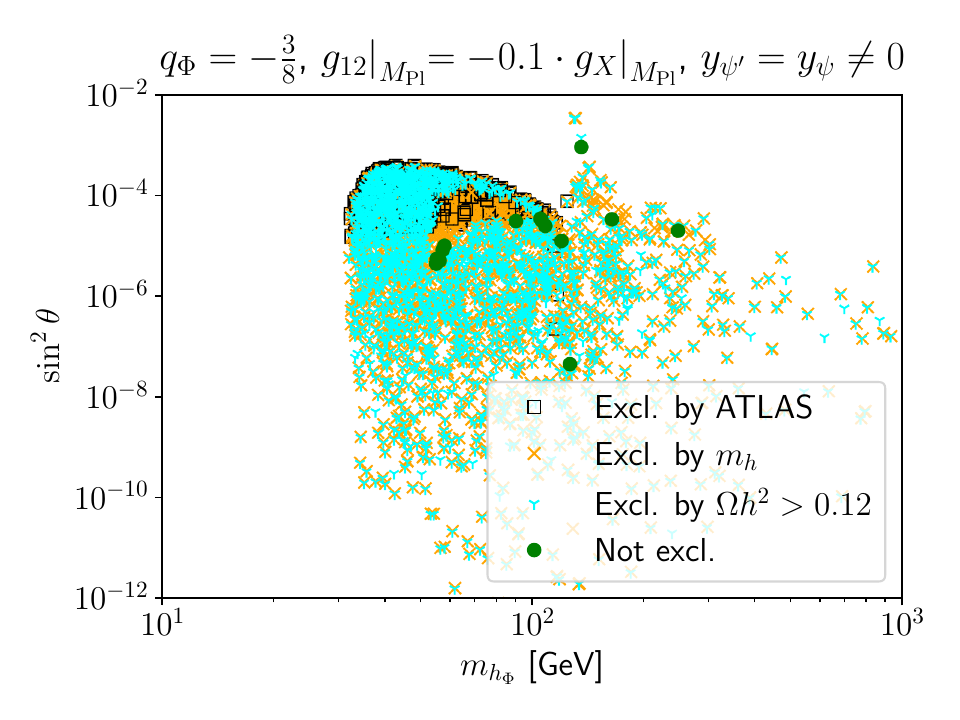}%
	\caption{Numerical values for the Higgs-dilaton mixing angle $\theta$ as a function of the dilaton mass $m_{h_\Phi}$ for the minimal model (top left), the neutrino portal model with $y_\psi\neq0$, $g_{12}\bigr|_{M_\mathrm{Pl}}=0$ (top right) and with $y_\psi\neq0$, $g_{12}\bigr|_{M_\mathrm{Pl}}=0$ (bottom left) and for the DM model with $g_{12}\bigr|_{M_\mathrm{Pl}}=0$ (bottom middle) and with $g_{12}\bigr|_{M_\mathrm{Pl}}=-0.1\cdot g_X\bigr|_{M_\mathrm{Pl}}$ (bottom right). The black squares indicate points excluded by the ATLAS dilepton resonance searches and the orange crosses indicate points where the Higgs mass is outside of its $3\sigma$ range. The green points are not excluded. For the DM model, the cyan ``tri-down'' points indicate the region where the relic abundance is $\Omega h^2>0.12$.}
	\label{fig. theta}
\end{figure}

Numerical values for the Higgs-dilaton mixing angle are shown in Fig.~\ref{fig. theta}. The mixing is suppressed by the heavy scale (see Eq.~(\ref{eq. theta Higgs dilaton})) and for points not excluded by ATLAS, the mixing is typically $\sin^2\theta\lesssim10^{-5}$ and, therefore, well below the direct experimental limits on the mixing angle~\cite{ATLAS:2015ciy,Robens:2015gla, Falkowski:2015iwa}. In the degenerate scenario $m_{h_\Phi}\approx m_h$ the mixing can be larger (see also Eq.~(\ref{eq. theta Higgs dilaton})). 
The couplings of the dilaton to the SM induced by mixing are obtained by SM operators containing $h_\Phi$ rather than $h$, i.e.\ $\mathcal{O}_{h_\Phi}\approx\sin\theta\times\mathcal{O}^\mathrm{SM}_{h\to h_\Phi}$.
Additional couplings of the dilaton to pairs of gauge bosons originate from the trace anomaly, but are suppressed by $h_\Phi/v_\Phi$~\cite{Goldberger:2007zk,Chacko:2012sy,Bellazzini:2012vz}.
Generic constraints on dilatons, hence, are avoided due to the large value of $v_\Phi$~\cite{Ahmed:2015uqt,Ahmed:2019csf}, and the dilaton decay branching ratios are, to a good approximation, those of a SM Higgs with the mass of the dilaton. We estimate that, at a Higgs factory, for $\sin^2\theta\sim\mathcal{O}(10^{-5})$, there would roughly be one dilaton produced per $10^5$ Higgs bosons and decay with a lifetime of $\tau_{h_\Phi\rightarrow\mathrm{SM}}\sim\mathcal{O}(10^{-17}\,\mathrm{s})$. While a smaller mixing angle would further decrease the dilaton production yield, it would also lead to an increased dilaton lifetime.
For long enough dilaton lifetimes, this would open a very prominent signature in displaced vertex tracking at Higgs factories. 
For example, for $\sin^2\theta\sim\mathcal{O}(10^{-7})$, $10^7$ Higgs bosons would be enough to yield a displaced vertex signature for a dilaton decay if $\mathcal{O}(\mu\mathrm{m})$ vertex tracker resolution could be achieved~\cite{ILDConceptGroup:2020sfq} (see also \cite{Ripellino:2024tqm}). Searches of this kind would also benefit from the primary vertex boost inherent to recently proposed asymmetrical beam configurations called HALHF~\cite{Foster:2023bmq,Laudrain:2023zaa}.

There would also be additional rare decays of the Higgs boson, or ``dilaton strahlung''\footnote{
AT is grateful to Ian M.\ Lewis for drawing our attention to this process.
} emitted from virtual Higgses. The three-scalar vertices, approximated for small custodial symmetry violation ($g_{12}\ll1$, $y_\psi\ll1$ and $\lambda_\Phi-\lambda_p\ll 1$), are given by
\begin{align}
	\frac{\partial^3 V_\text{eff}}{\partial h^3}&\approx6\lambda_H v_H,\\
	\frac{\partial^3 V_\text{eff}}{\partial h^2\partial h_\Phi}&\approx\left(m_{h_\Phi}^2-m_h^2\right)\frac{1}{v_\Phi},\\
	\frac{\partial^3 V_\text{eff}}{\partial h\partial h_\Phi^2}&\approx\left(3m_{h_\Phi}^2-m_h^2\right)\frac{v_H}{v_\Phi^2},
\end{align}
implying that Higgs decays into dilatons, if kinematically allowed, are highly suppressed with a branching fraction of $\Gamma_{h\to h_\Phi h_\Phi}/\Gamma_{h,\mathrm{tot}}\sim\mathcal{O}(10^{-8})$ and also dilaton strahlung is suppressed by the small mixing angle. 

Since Custodial Naturalness is based on new sources of custodial symmetry violation, also electroweak precision tests (EWPT) provide meaningful constraints on our models. 
To a first approximation, the new sources of custodial breaking induce a shift of the mass of the $Z$ boson visible in Eq.~\eqref{eq:mZmZp}. If all other couplings keep their SM values, $m_Z$ would stay within its $2\sigma$ uncertainty~\cite{ParticleDataGroup:2022pth} if $\langle\Phi\rangle\gtrsim18\,\mathrm{TeV}$. This constraint is always superseded by 
direct limits on the $Z'$ mass, see Figs.~\ref{fig. minimal model}, \ref{fig. fid cross}, and \ref{fig. neutrino model}, which justifies our simplistic treatment here. For a detailed analysis, a new global fit to the wealth of EWPT data would be in order,
since it can also explore additional parameter correlations in our class of models. This is, however, beyond the scope of this paper.

\begin{figure}
	\centering
	\includegraphics[width=0.5\textwidth]{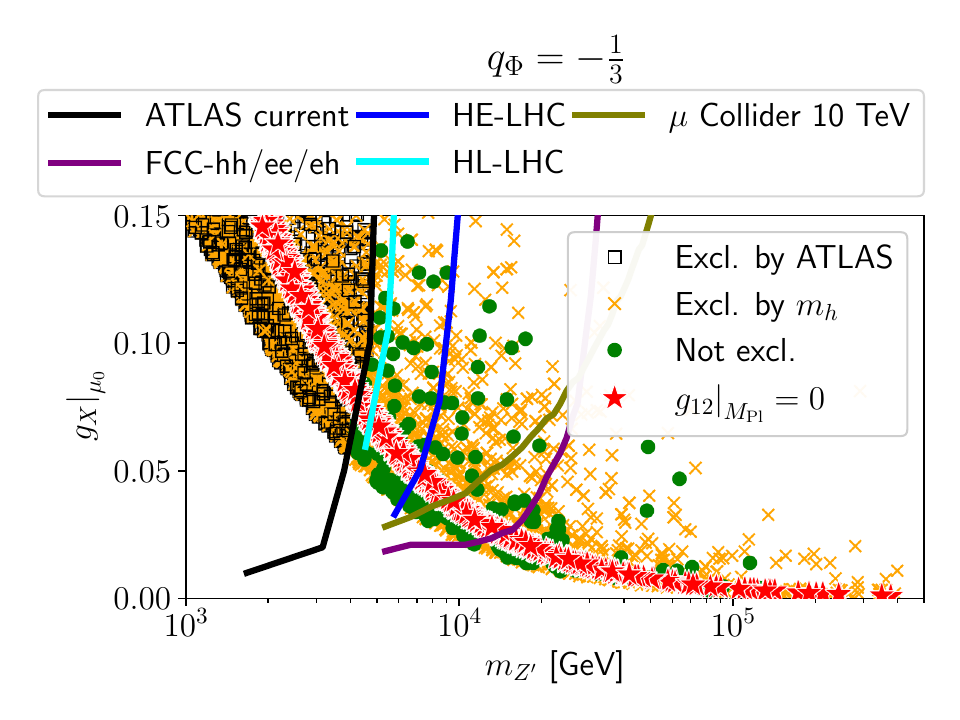}%
	\includegraphics[width=0.5\textwidth]{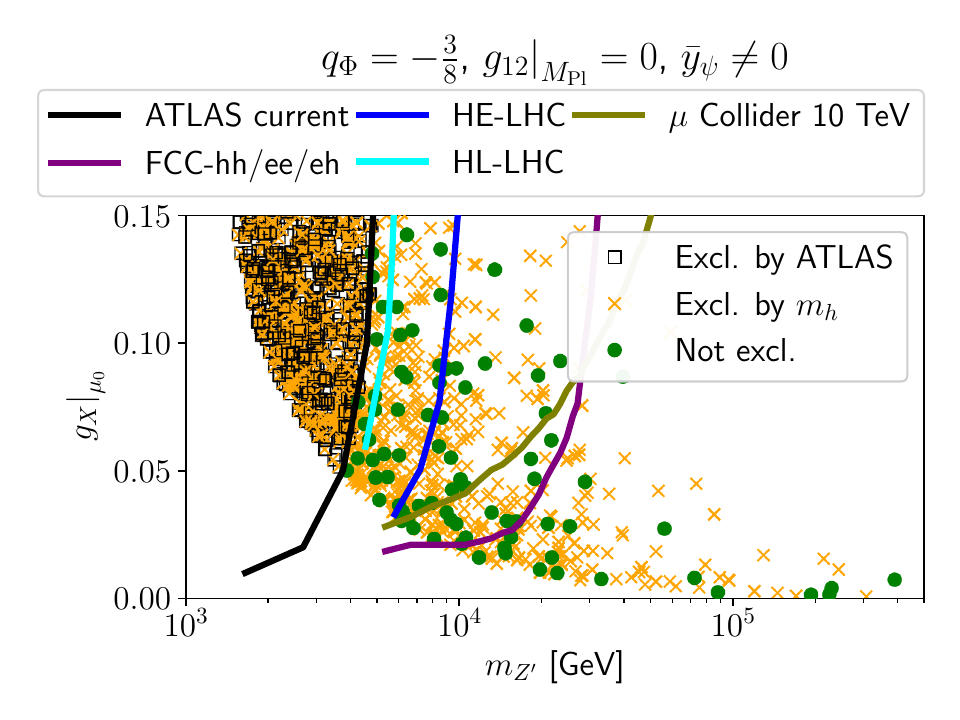}
	\includegraphics[width=0.33\textwidth]{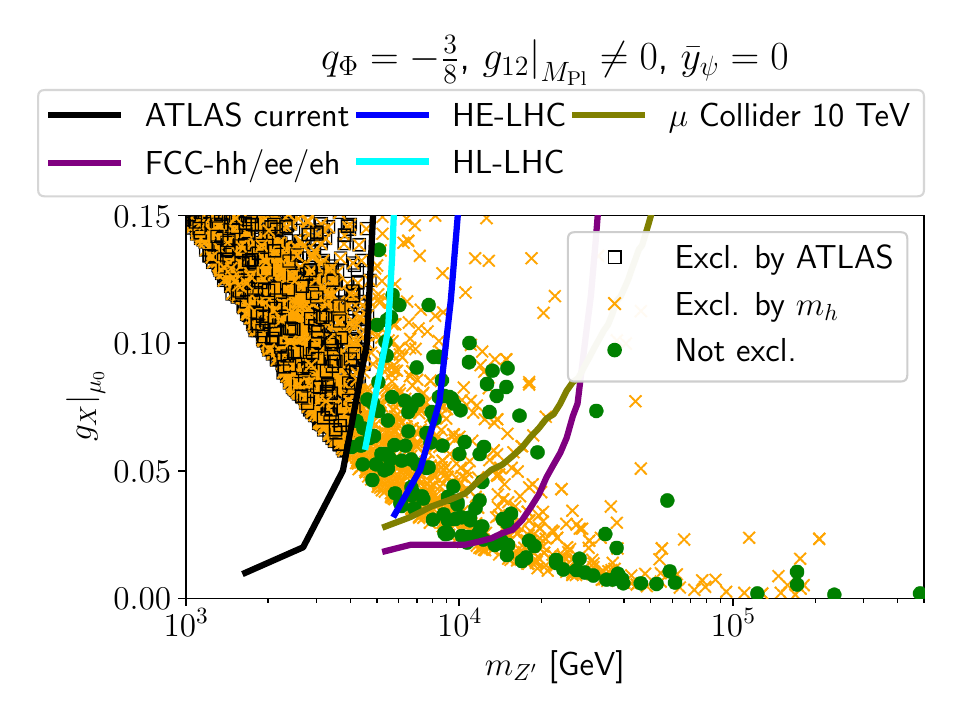}%
	\includegraphics[width=0.33\textwidth]{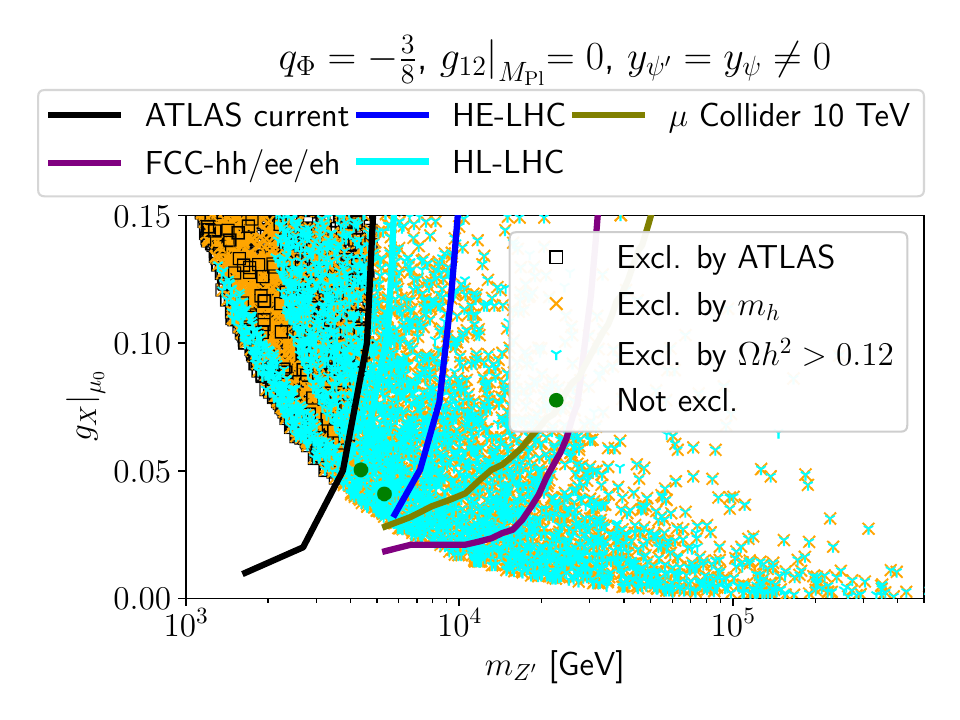}%
	\includegraphics[width=0.33\textwidth]{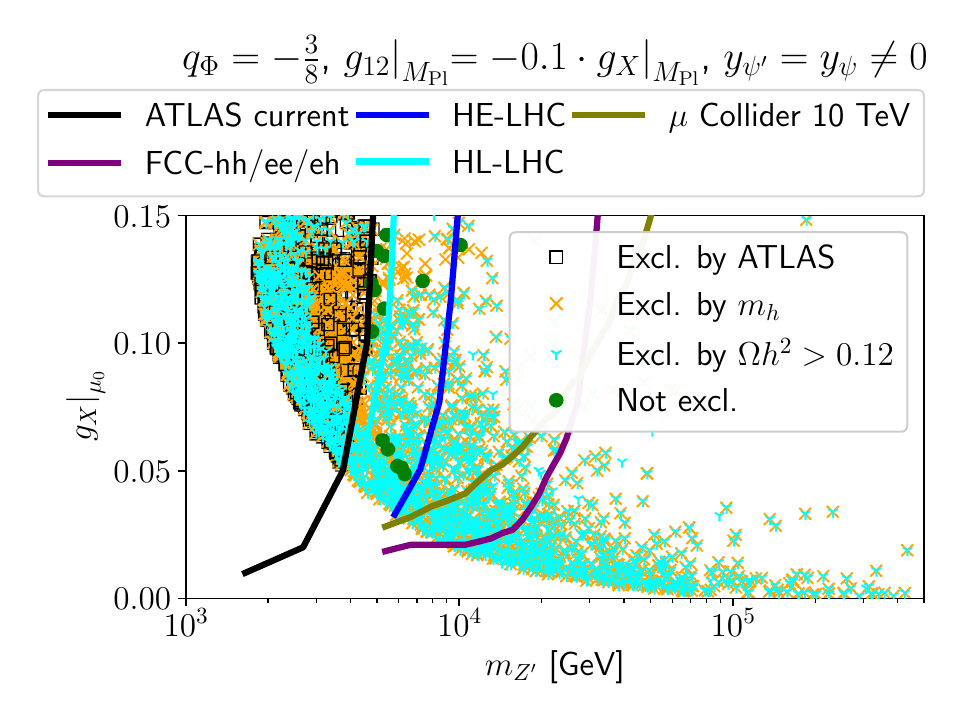}%
 	\caption{$\U{1}_\mathrm{X}$ gauge coupling at the matching scale $\mu_0$ vs.\ $m_{Z'}$ for the minimal model (top left), the neutrino portal model with $y_\psi\neq0$, $g_{12}\bigr|_{M_\mathrm{Pl}}=0$ (top right) and with $y_\psi\neq0$, $g_{12}\bigr|_{M_\mathrm{Pl}}=0$ (bottom left) and for the DM model with $g_{12}\bigr|_{M_\mathrm{Pl}}=0$ (bottom middle) and with $g_{12}\bigr|_{M_\mathrm{Pl}}=-0.1\cdot g_X\bigr|_{M_\mathrm{Pl}}$ (bottom right). The black squares indicate points excluded by the ATLAS dilepton resonance searches and the orange crosses indicate points where the Higgs mass is outside of its $3\sigma$ range~\cite{ParticleDataGroup:2022pth}. The green points are not excluded. For the DM model, the  cyan ``tri-down'' points indicate the region where the relic abundance is $\Omega h^2>0.12$. We also show the recast ATLAS limits~\cite{ATLAS:2019erb} and projections for future colliders taken from Ref.~\cite{EuropeanStrategyforParticlePhysicsPreparatoryGroup:2019qin}. The projections assume a hypercharge universal $Z'$.}
	\label{fig. Zp limit}
\end{figure}

Finally, The $Z'$ boson can be searched for at future colliders and we show the reach of different proposals in Fig.~\ref{fig. Zp limit}. The projected future limits are taken from Ref.~\cite[Fig.~8.3]{EuropeanStrategyforParticlePhysicsPreparatoryGroup:2019qin} but have been calculated for a hypercharge universal $Z'$. A more detailed analysis can be done, taking into account the non-universal 
$Z'$ couplings in our model but results do not vary by more than an $\mathcal{O}(1)$ factor. For the DM model, all points found in our scans can be excluded by $Z'$ searches at future colliders. Smaller values of $g_{12}\bigr|_{M_\mathrm{Pl}}$ than the ones considered here help to escape such searches but would require fine tuning of gauge kinetic mixing  against Yukawa couplings.

\section{Cosmological evolution and gravitational wave signatures}\label{sec. finite T}
In this work we have only considered the zero temperature effective potential. Future work should investigate finite temperature effects in the class of Custodial Naturalness models. Coleman-Weinberg type models generically have a first order phase transition (FOPT)~\cite{Linde:1975sw,Weinberg:1976pe,Litim:1994jd}.
The thermal history of the classically conformal $B-L$ model has been studied for example in Refs.~\cite{Jinno:2016knw,Iso:2017uuu,Marzo:2018nov,Ellis:2020nnr,Schmitt:2024pby} and connections to potentially realistic scenarios of Baryo- or Leptogenesis (sometimes  also including the production of DM) have been made in Refs.~\cite{Iso:2010mv,Konstandin:2011dr,Khoze:2013oga,Hambye:2018qjv,Huang:2022vkf,Dasgupta:2022isg,Das:2024gua,Chauhan:2024jfq}. We briefly summarize the main findings. The conformal $B-L$ model differs from our model by the usually considered charge assignment $q_\Phi=2$ and the fact that \SO{6} custodial symmetry is not realized. The different charge of $\Phi$ can largely be compensated by rescaling the gauge coupling. Note that the qualitative behavior also holds for different values of gauge kinetic mixing~\cite{Marzo:2018nov}.

At sufficiently high temperature $T$, the minimum of the potential is $(\Phi,H)=(0,0)$. Once the temperature drops, the potential develops a non-trivial minimum. Below the critical temperature $T_c\sim m_{Z'}$, this non-trivial minimum has lower energy than the false vacuum $\Phi=0$. As a consequence of classical scale invariance, there is always a thermal barrier which does not disappear at low temperature and $\Phi$ remains trapped at $\Phi=0$. The field $\Phi$ will tunnel to the non-trivial vacuum leading to the formation of bubbles. However, the percolation temperature $T_p$ where the formation and expansion of bubbles becomes efficient is much lower than the critical temperature $T_p\ll T_c$. This leads to a period of thermal inflation and the number of $e$-folds for the parameters of our model is typically $N\sim 10$~\cite{Marzo:2018nov}.
It turns out that for large parts of the parameter space, percolation remains inefficient even for temperatures below the QCD scale. 
Ref.~\cite{Iso:2017uuu} found that in the $B-L$ model this happens if $g_\mathrm{B-L}\lesssim0.2$ at $\mu=m_{Z'}$. Rescaling this number as the conformal $B-L$ model corresponds to $q_\Phi=2$, we find that for our model this bound translates to $g_X\lesssim 0.4$. In this case, the QCD phase transition occurs before $B-L$ symmetry breaking. 
QCD with $N_f=6$ massless quarks has a FOPT with a critical temperature $T^\mathrm{QCD}_c\approx 85\,\mathrm{MeV}$~\cite{Braun:2006jd,Cuteri:2021ikv}.
The top quark condensate generates a linear term in the Higgs potential which in turn induces a VEV for the Higgs boson given by $v_{H,\mathrm{QCD}}=\left|y_t/(\sqrt2 \lambda_H)\langle t\bar t\rangle\right|^{1/3}\approx 100\,\mathrm{MeV}$~\cite{Iso:2017uuu,Schmitt:2024pby}.
For the parameter space of our model, Ref.~\cite{Iso:2017uuu} finds that $\Phi$ initially remains trapped at $\Phi=0$ and transitions to the true vacuum by a FOPT. Ref.~\cite{Schmitt:2024pby} suggests that a  QCD induced tachyonic instability leads to $B-L$ breaking without another FOPT. In both cases, bubble collisions lead to gravitational wave signal and parts of the parameter space can be probed by future gravitational wave observatories~\cite{Jinno:2016knw,Marzola:2017jzl,Iso:2017uuu,Prokopec:2018tnq,Marzo:2018nov,Ellis:2020nnr,Dasgupta:2022isg,Huang:2022vkf,Sagunski:2023ynd,Schmitt:2024pby}.

Future work also should investigate the reheating process. If $m_{h_\Phi}>2m_h$, which only holds in a small part of our parameter space, then the decay $h_\Phi\to h h$ reheats the thermal bath and the reheating temperature $T_\mathrm{rh}\sim\mathcal{O}(\mathrm{TeV})$~\cite{Schmitt:2024pby}. For most of our parameter space, $m_{h_\Phi}<2m_h$. Reheating might still be possible, for example through scalar mixing~\cite{Kawai:2023dac}. 

Altogether this shows that the cosmological history of our model can be realistic, and that the model can be probed by its gravitational wave signal. More detailed investigations are necessary, however, in order to work out reliable quantitative predictions for the specific class of models realizing Custodial Naturalness.

The models considered here introduce three light fermionic degrees of freedom in form of the right handed components of the Dirac neutrinos. In the minimal case and in the DM model we have three Dirac neutrinos, while in the neutrino portal model we have two Dirac neutrinos and the massless Weyl spinor $\psi_R$. These new light degrees of freedom can be produced via the $Z'$ portal and contribute to $\Delta N_\text{eff}$. Details depend on the production of these degrees of freedom after thermal inflation. Assuming complete thermalization, the contribution would be given by $\Delta N_\text{eff}\approx0.14$, which can be probed by future CMB experiments~\cite{Heeck:2014zfa,Abazajian:2019oqj,Adshead:2022ovo}.

\section{Variations and embeddings of Custodial Naturalness}\label{sec. future directions}
We have shown that the most minimal version of Custodial Naturalness~\cite{deBoer:2024jne} is phenomenologically stable under variations of high-scale boundary conditions and can easily be extended by additional new fermions to incorporate neutrino mass generation and/or fermionic DM candidates. In the present study, we have focused on what we think is the most interesting region of parameter space where custodial symmetry is restored around $\mu\sim M_\mathrm{Pl}$. Similar to the SM, this class of  models feature a Higgs vacuum in- or better meta-stability at scales around $\mu\sim10^{15}-10^{17}\,\mathrm{GeV}$, which could be avoided by the introduction of additional fermions (see e.g.\ Refs.~\cite{Oda:2015gna,Das:2015nwk,Das:2016zue}). On the other hand, 
vacuum meta-stability may not be something that needs to be ``cured'' but could also be an important feature of Nature like in the scenario of multi-phase criticality~\cite{Froggatt:1995rt} (see also~\cite{Kannike:2021iyh,Huitu:2022fcw,Kannike:2022pva}), or if scale generation and separation is related to an interacting UV fixed point (see e.g.\ Ref.~\cite{Litim:2014uca}) with custodially-symmetric quantum critical values of the scalar self-couplings.

A totally different but likewise valid possibility is that the scale of custodial symmetry violation is lowered to $\mu\approx 10^{11}\,\mathrm{GeV}$, as already remarked in footnote~\ref{footnote lower CS}.
If the scale of custodial symmetry violation is reduced, then SM contributions to $\beta_{\lambda_p}-\beta_{\lambda_\Phi}$ can be sufficiently large to trigger EWSB without requiring additional sources of custodial symmetry breaking. Specifically, for the charge assignment $q_\Phi=-\frac{16}{41}$, $g_{12}$ remains zero at one loop and does not contribute to custodial symmetry violation. 
If we assume that a new high scale can be generated in a scale invariant setting independently from the low scale, then special charge assignments could be justified if the $\U{1}_{\mathrm{X}}$ together with the SM gauge group could be embedded into a larger simple group similar to a grand unified theory (GUT). The enhanced \SO{6} custodial symmetry then might be embedded similarly to the SM custodial symmetry $\SO{4}\subset\SO{10}$ in Pati-Salam unification~\cite{Pati:1974yy}.

While we have exclusively considered family universal $\U{1}_\mathrm{X}$ assignments in this work, extensions of the idea of Custodial Naturalness to non-family-universal charge assignments could extend our mechanism of scale separation to the flavor structure of the SM. Extensions such as $B-L$ with charge assignment of right-handed neutrinos $\nu_R\sim(-4,-4,5)$ could explain the smallness of neutrino masses~\cite{Montero:2007cd,Ma:2014qra,Ma:2015mjd,Bonilla:2018ynb}, while 
a charge assignment consistent with $\U{1}_{\mathrm{L}_\mu-\mathrm{L}_\tau}$ might allow an explanation of the observed discrepancy in the muon anomalous magnetic moment~\cite{He:1990pn,Baek:2001kca,Ma:2001md,Heeck:2011wj,Altmannshofer:2014pba,Altmannshofer:2016oaq}.

Bosonic contributions to the running of $\lambda_\Phi$ are required in order for $\lambda_\Phi$ to reach critical values. In the class of models discussed here, these contribution come from the $\U{1}_X$ gauge boson. An alternative possibility would be to introduce an additional real scalar field $S$ instead of the $\U{1}_X$ gauge group. In this case, also $\Phi$ can be constrained to be a real scalar field. The custodial symmetry of the $H-\Phi$ system then would be given by $\SO{5}$, again leading to the correct number of pNGB's to cover the dilaton and SM-like Higgs field. The field $S$ in this scenario would be a singlet under the enhanced custodial symmetry. Since $S$ can be charged under an additional $\mathbbm{Z}_2$ symmetry, it would be a good dark matter candidate similar to the scenario considered in Ref.~\cite{Kannike:2022pva}. In this scenario, Yukawa interactions involving right handed neutrinos and $\Phi$ generate Majorana mass terms. The left handed neutrinos then obtain masses via a type I seesaw mechanism. Realistic neutrino masses can be obtained with (Dirac) Yukawa couplings $y_\nu\sim10^{-5}$.

Finally, it may also be interesting to investigate scenarios in which spontaneous scale generation is linked to inflation~\cite{Kubo:2020fdd,Kubo:2018kho}, where in our case $\Phi$ would have to play the role of the inflaton, see e.g.~Refs.~\cite{Okada:2013vxa,Oda:2017zul,Kawai:2023dac}.

\section{Conclusions}\label{sec. Conclusion}
We have introduced Custodial Naturalness which is a new mechanism to address the hierarchy problem. The large separation between the Planck scale and a new intermediate scale is generated via dimensional transmutation. The further suppression of the EW scale is naturally explained by the fact that the Higgs boson is a pNGB of an enhanced custodial symmetry that is spontaneously broken at the intermediate scale.

The scalar sector consist of the SM Higgs field and a complex scalar singlet, both of which have identical charges under a new gauged \U{1} symmetry. At some high scale, which we take to be the Planck scale, the potential is assumed to be scale invariant and invariant under a \SO{6} custodial symmetry. An intermediate scale is generated via the Coleman-Weinberg mechanism, spontaneously breaking scale and custodial symmetry. The Higgs boson is identified as a pNGB associated with the spontaneous breaking of custodial symmetry, therefore avoiding the little hierarchy problem.
In our analytical discussion we investigated the impact of explicit custodial symmetry breaking on the Higgs mass. The leading contributions are found to come from gauge kinetic mixing and the Yukawa couplings of potential new fermions while the SM gauge and Yukawa couplings only contribute in a subleading manner.

The minimal realization of Custodial Naturalness consists of the SM extended by a complex scalar singlet and a new $\U{1}_\mathrm{X}$ gauge symmetry. With the boundary condition of vanishing gauge kinetic mixing at the Planck scale, this model is predictive because it has the same number of parameters as the SM. We have shown that the Custodial Naturalness mechanism is stable under the inclusion of additional fields and new sources of custodial symmetry violation. The minimal fermionic extension connects the new sector to the neutrino portal and predicts two massive Dirac neutrinos, an exactly massless lightest active neutrino, as well as a heavy sterile Dirac neutrino. We also presented a model which naturally encompasses two-component DM and the relic abundance reaches the observed value in a small part of the parameter space which requires some tuning between mass of the heavy vector boson $Z'$ mediator and the DM fermion masses.
Both models allow for new Yukawa interactions which violate custodial symmetry and, therefore, indirectly contribute to the Higgs mass. For each model we demonstrated how custodial symmetry violation impacts the hierarchy between the EW and the intermediate scale and we showed, using a variation of the Barbieri-Giudice measure, that our mechanism does not require fine tuning. 

The realizations of Custodial Naturalness considered here predict a heavy $Z'$ boson in the $4-100\,\mathrm{TeV}$ mass range with couplings to all SM fields. Future colliders can probe large parts of the allowed parameter space. If there is no additional tuning, the entire parameter space of the DM model can be tested by future $Z'$ searches.
The Dilaton, which is the pNGB associated with spontaneous breaking of scale symmetry, typically has a mass in the  $30-1000\,\mathrm{GeV}$ range, small mixing with the SM Higgs boson, and potentially long enough lifetime to provide a benchmark case for displaced vertex searches at future Higgs factories.
For most of the parameter space, the top pole mass is required to be at the lower end of its currently experimentally allowed $1\sigma$ range. 

The thermal history of the Universe for models similar to ours (i.e.\ with scale invariance and a similar particle content) has been studied in previous works. Such settings typically feature a strongly supercooled first-order phase transition which gives rise to potentially observable gravitational wave signals. Details of the cosmological evolution in our models and the implications for gravitational wave observatories should be investigated in future work.

Variations of our minimal scenarios can connect the idea of Custodial Naturalness to the flavor structure of the SM while embeddings in unified theories can further constrain the possible charge assignments and may provide insights about the origin of high scale custodial symmetry.

\section*{Acknowledgments}
The work of AT was partially supported by the Portuguese Funda\c{c}\~ao para a Ci\^encia e a Tecnologia (FCT) through project 2023.06787.CEECIND and contract \href{https://doi.org/10.54499/2024.01362.CERN}{2024.01362.CERN}, partially funded through POCTI (FEDER), COMPETE, QREN, PRR, and the EU.

\bibliographystyle{JHEP}
\bibliography{bib.bib}

\end{document}